%% file: main.tex
\documentclass[sigconf]{aamas}

\usepackage{balance} % for balancing columns on the final page

\usepackage{amsmath}
\usepackage{algorithm}
\usepackage{algpseudocode}
\usepackage{balance}

\usepackage{booktabs, multirow} % for borders and merged ranges
\usepackage{soul}% for underlines
\usepackage{xcolor,colortbl} % for cell colors
\usepackage{changepage,threeparttable} % for wide tables
\usepackage{graphicx}

\usepackage{subcaption}  
\usepackage{float}       
\usepackage{caption}

\usepackage{bbm}

\makeatletter
\gdef\@copyrightpermission{
  \begin{minipage}{0.2\columnwidth}
   \href{https://creativecommons.org/licenses/by/4.0/}{\includegraphics[width=0.90\textwidth]{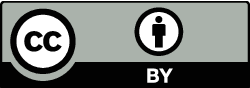}}
  \end{minipage}\hfill
  \begin{minipage}{0.8\columnwidth}
   \href{https://creativecommons.org/licenses/by/4.0/}{This work is licensed under a Creative Commons Attribution International 4.0 License.}
  \end{minipage}
  \vspace{5pt}
}
\makeatother

\setcopyright{ifaamas}
\acmConference[AAMAS '25]{Proc.\@ of the 24th International Conference
on Autonomous Agents and Multiagent Systems (AAMAS 2025)}{May 19 -- 23, 2025}
{Detroit, Michigan, USA}{Y.~Vorobeychik, S.~Das, A.~Nowé  (eds.)}
\copyrightyear{2025}
\acmYear{2025}
\acmDOI{}
\acmPrice{}
\acmISBN{}

\acmSubmissionID{533}

\title[AAMAS-2025 Formatting Instructions]{Dynamic Sight Range Selection in Multi-Agent Reinforcement Learning}

\author{Wei-Chen Liao}
\affiliation{
  \institution{Department of Computer Science, National Yang Ming Chiao Tung University}
  \city{Hsinchu}
  \country{Taiwan}}
\email{wcl.cs11@nycu.edu.tw}

\author{Ti-Rong Wu}
\affiliation{
  \institution{Institute of Information Science, Academia Sinica}
  \city{Taipei}
  \country{Taiwan}}
\email{tirongwu@iis.sinica.edu.tw}

\author{I-Chen Wu}
\affiliation{
  \institution{Department of Computer Science, National Yang Ming Chiao Tung University}
  \city{Hsinchu}
  \country{Taiwan}}
\email{icwu@cs.nycu.edu.tw (correspondence)}

\begin{abstract}
Multi-agent reinforcement Learning (MARL) is often challenged by the sight range dilemma, where agents either receive insufficient or excessive information from their environment.
In this paper, we propose a novel method, called \textit{Dynamic Sight Range Selection (DSR)}, to address this issue.
DSR utilizes an Upper Confidence Bound (UCB) algorithm and dynamically adjusts the sight range during training.
Experiment results show several advantages of using DSR.
First, we demonstrate using DSR achieves better performance in three common MARL environments, including Level-Based Foraging (LBF), Multi-Robot Warehouse (RWARE), and StarCraft Multi-Agent Challenge (SMAC).
Second, our results show that DSR consistently improves performance across multiple MARL algorithms, including QMIX and MAPPO.
Third, DSR offers suitable sight ranges for different training steps, thereby accelerating the training process.
Finally, DSR provides additional interpretability by indicating the optimal sight range used during training.
Unlike existing methods that rely on global information or communication mechanisms, our approach operates solely based on the individual sight ranges of agents.
This approach offers a practical and efficient solution to the sight range dilemma, making it broadly applicable to real-world complex environments.
\end{abstract}


\newcommand{\BibTeX}{\rm B\kern-.05em{\sc i\kern-.025em b}\kern-.08em\TeX}

\begin{document}

%%% The following commands remove the headers in your paper. For final 
%%% papers will be inserted during the pagination process.

\pagestyle{fancy}
\fancyhead{}

%%% The next command prints the information defined in the preamble.

\maketitle 

%%%%%%%%%%%%%%%%%%%%%%%%%%%%%%%%%%%%%%%%%%%%%%%%%%%%%%%%%%%%%%%%%%%%%%%%

% \input{main}
\section{Introduction}\label{sec:intro}
Reinforcement learning \cite{sutton_ReinforcementLearning_2018} has achieved significant success in various domains, such as gaming \cite{schrittwieser_MasteringAtari_2020}, circuit design \cite{mirhoseini_GraphPlacement_2021}, and recommendation systems \cite{afsar_ReinforcementLearning_2022}.
To extend its applicability to complex real-world problems, particularly those involving multiple agents that must cooperate or compete to achieve shared goals, multi-agent reinforcement learning (MARL) \cite{ning_SurveyMultiagent_2024} has recently emerged to address the challenges in multi-agent environments, including multiplayer gaming \cite{vinyals_GrandmasterLevel_2019}, autonomous vehicles \cite{zhang_MultiAgentReinforcement_2024}, robotic control \cite{xiao_AsynchronousActorCritic_2022}, and traffic signal control \cite{zhao_SurveyDeep_2024, noaeen_ReinforcementLearning_2022}.

Various approaches have been developed in cooperative MARL.
For instance, independent learning (IL) techniques, such as Independent Q-Learning (IQL) and Independent Proximal Policy Optimization (IPPO) \cite{schulman_ProximalPolicy_2017, dewitt_IndependentLearning_2020}, train each agent individually, treating other agents as part of the environment.
While this simplifies the learning process by reducing the complexity of considering all agents simultaneously, it also introduces the challenge of non-stationarity, where the learning dynamics continuously change as other agents adapt during training \cite{yang_OverviewMultiAgent_2020}.
On the other hand, Centralized Training with Decentralized Execution (CTDE) has been a widely used framework for cooperative MARL problems, addressing non-stationarity issues by allowing agents to access all available information during centralized training while maintaining decentralized decision-making.
Successful methods based on CTDE include Q-learning-based approaches like QMIX \cite{rashid_QMIXMonotonic_2018}, QTRAN \cite{son_QTRANLearning_2019}, and QPLEX \cite{wang_QPLEXDuplex_2021}, as well as policy optimization methods like MAPPO \cite{yu_SurprisingEffectiveness_2022} and HAPPO \cite{kuba_TrustRegion_2022}.

\begin{figure}[htb] 
  \centering
  \begin{subcaptionblock}{0.32\columnwidth}
    \includegraphics[width=\textwidth]{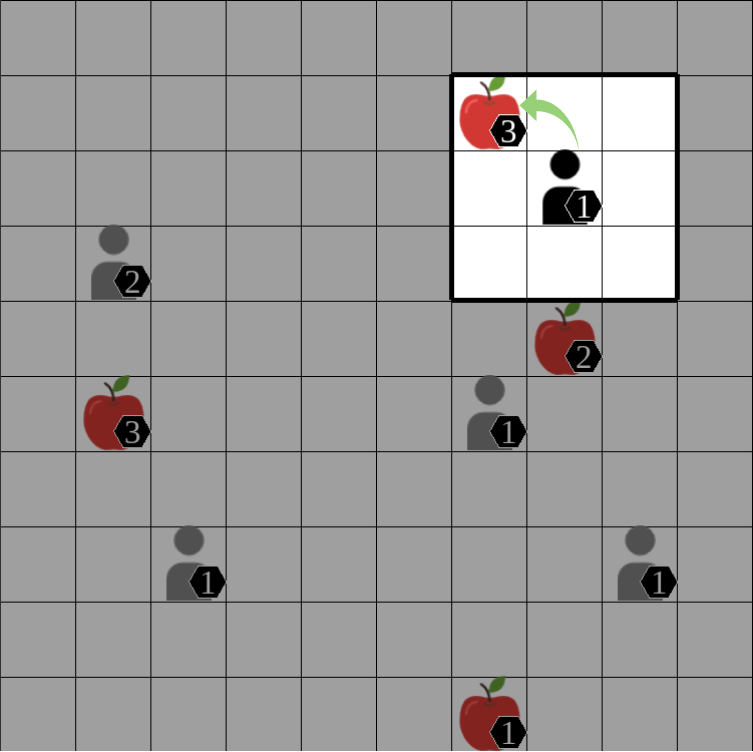}
    \caption{}
    \label{fig:sight_range_dilemma_1s}
  \end{subcaptionblock}
  \hfill
  \begin{subcaptionblock}{0.32\columnwidth}
    \includegraphics[width=\textwidth]{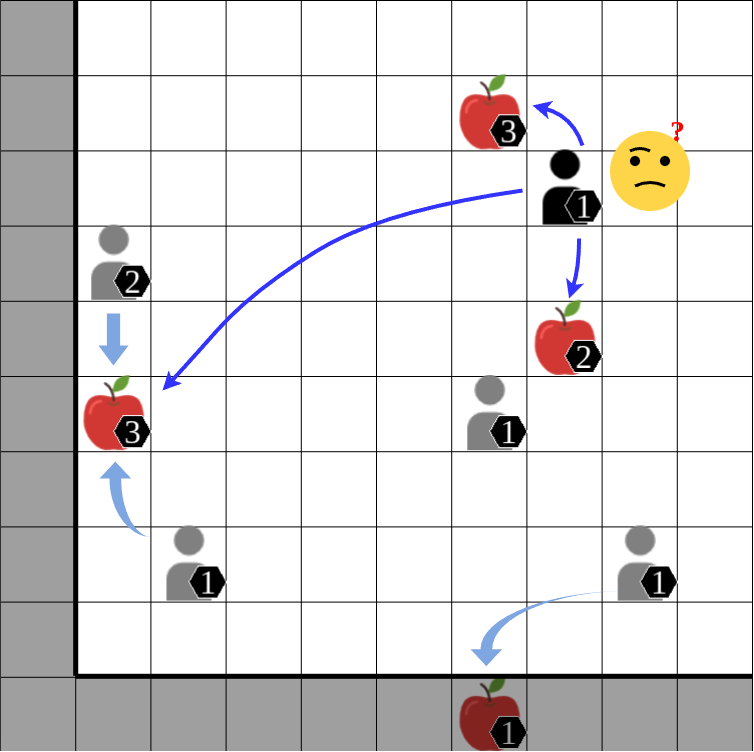}
    \caption{}
    \label{fig:sight_range_dilemma_6s}
  \end{subcaptionblock}
  \hfill
  \begin{subcaptionblock}{0.32\columnwidth}
    \includegraphics[width=\textwidth]{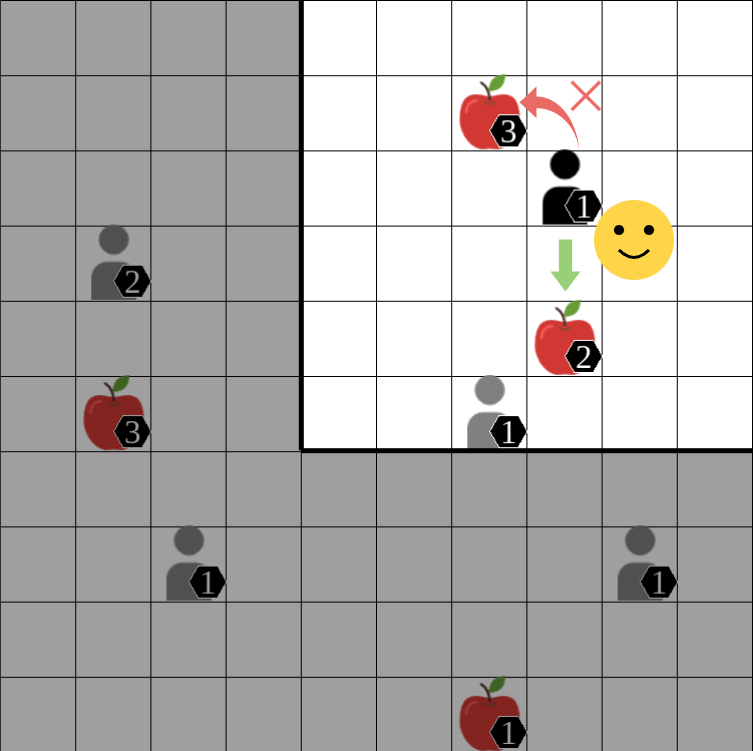}
    \caption{}
    \label{fig:sight_range_dilemma_3s}
  \end{subcaptionblock}
  \caption{An illustration of the sight range dilemma in the level-based foraging game, where players are required to cooperate to collect food. (a) With a small sight range, the player may fail to see other players. (b) With a large sight range, the player receives excessive information irrelevant to their decision. (c) With an appropriate sight range, the player can easily identify the right partner to cooperate with in collecting food.}
  \label{fig:sight_range_dilemma}
\end{figure}

Due to the partial observability in MARL, one of the key challenges is limited information that each agent can observe from the environment, often referred to as the \textit{sight range}.
In real-world applications, agents usually perceive only a limited portion of the environment due to constraints such as sensor range.
For instance, in autonomous driving, each vehicle can detect only a small region of its surroundings rather than the entire environment.
A small sight range generally provides insufficient information, making it difficult for agents to make effective decisions.
However, previous research suggests that a larger sight range is not always better, as it often includes excessive and irrelevant information, which hinders the learning process and leads to worse performance \cite{guan_EfficientMultiagent_2022,shao_ComplementaryAttention_2023}.
The trade-off in selecting an appropriate sight range, known as the \textit{sight range dilemma} \cite{shao_ComplementaryAttention_2023}, remains a critical challenge in MARL, as illustrated in Figure \ref{fig:sight_range_dilemma}.
Previous works \cite{shao_ComplementaryAttention_2023, guan_EfficientMultiagent_2022} have addressed this dilemma, mainly focusing on communication mechanisms that use self-attention to leverage global information, enabling each agent to identify relevant agents and adjust its sight range accordingly.
Although these methods mitigate the sight range dilemma, they rely on acquiring global information for all agents, which is often impractical in real-world environments.

To tackle this challenge, we propose \textbf{Dynamic Sight Range Selection (DSR)}, a novel approach that directly addresses the sight range dilemma without requiring global information.
Specifically, DSR dynamically adjusts the sight range during training using an Upper Confidence Bound (UCB) algorithm \cite{garivier_UpperConfidenceBound_2008}, allowing the agent to be trained with varying sight ranges and converge on the most suitable sight range.
Unlike previous methods that rely on global information, DSR controls only the sight ranges of individual agents, offering clearer insights into how much information is needed in different environments.
Moreover, DSR can automatically discover the optimal sight range, which is particularly valuable for real-world applications.
For instance, in autonomous systems, the sight range obtained through DSR can guide sensor design, helping to balance sight range, reduce costs, and maintain high performance.

The contributions of this paper are summarized as follows:
\begin{itemize}
    \item DSR effectively addresses the sight range dilemma issue and outperforms the baseline model without DSR in three common MARL environments, including LBF, RWARE, and SMAC.
    \item By training agents with different sight ranges, DSR also accelerates the training process.
    \item DSR can be simply integrated with any MARL algorithm such as QMIX and MAPPO.
    \item The dynamically selected sight ranges further provide additional interpretability and explainability.
\end{itemize}

\section{Background}\label{sec:bg}

\subsection{Multi-agent Reinforcement Learning}\label{sec:bg-marl}

Cooperative multi-agent problems can be modeled as Decentralized Partially Observable Markov Decision Processes (Dec-POMDPs) \cite{bernstein_ComplexityDecentralized_2000, oliehoek_ConciseIntroduction_2016}, defined as $\mathcal{M} = \langle \mathcal{N}, \mathcal{S}, \mathcal{A}, \mathcal{P}, \mathcal{O}, \mathcal{Z}, \mathcal{R}, \gamma \rangle$, where $\mathcal{N} = \{n_1, n_2,\\ \dots, n_N\}$ is the set of agents, $\mathcal{S}$ is the set of global states, $\mathcal{A}$ is the set of joint individual actions from each agent, $\mathcal{P}$ is the state transition function, $\mathcal{O}$ is the set of observations, $\mathcal{Z}$ is the observation function, $\mathcal{R}$ is the reward function, and $\gamma \in [0, 1)$ is the discount factor for calculating future rewards.
Given a state $s^t$ at timestep $t$, where $s^t$ represents the global state with full observability, each agent $n_i \in \mathcal{N}$ receives a partial observation $o^t_i \in \mathcal{O}$, which is derived from the observation function $\mathcal{Z}(s^t, n_i)$.
The agent then selects an action $a_i \in \mathcal{A}$ based on its policy $\pi_i(a_i \mid \tau_i)$, where $\tau_i = \{ o_i^1, a_i^1, \dots, o_i^t \}$ is the action-observation history for agent $n_i$.
The joint action $a=\{a_1, a_2, \dots, a_N\}$ is formed by the actions selected by all agents, and the environment transitions to the next state $s^{t+1}$ according to the state transition function $\mathcal{P}(s^{t+1} \mid s^t, a)$.
All agents receive a shared reward $r^t$ from the reward function $\mathcal{R}(s^t, a)$ and obtain a new partial observation $o^{t+1}_i$.
This process repeats continuously until the termination.

The goal of Dec-POMDPs is to find the optimal joint policy of all agents, $\pi^*$, to maximize the global value function, $Q^{\pi}(s^t, a^t)=\mathbb{E} \left[ \sum_{t=0}^{\infty} \gamma^t R(s^t, a^t) \mid s_0 = s, \pi \right]$.
Due to the partial observability in Dec-POMDPs, $Q(\tau, a)$ is often used instead of $Q(s, a)$, where $\tau$ represents the history of observations and actions of all agents.
Namely, agents are supposed to make decisions based on incomplete information and learn an optimal policy $\pi_i$ that maximizes the total expected reward while cooperating with other agents in a partially observable environment.

\subsection{Centralized Training with Decentralized Execution (CTDE)}\label{sec:bg-ctde}
The \textit{Centralized Training with Decentralized Execution (CTDE)} framework has been widely adopted to address the challenges of partial observability in MARL \cite{sunehag_ValueDecompositionNetworks_2018, rashid_QMIXMonotonic_2018, wang_ROMAMultiAgent_2020, wang_QPLEXDuplex_2021, son_QTRANLearning_2019, yu_SurprisingEffectiveness_2022}.
CTDE utilizes centralized training, where agents have access to global states during the learning process.
However, during execution, decision-making is decentralized, and each agent relies only on its observation.
This approach allows agents to make independent decisions during execution while benefiting from coordinated training to enhance overall team performance.

In CTDE, value-based methods often employ the \textit{Individual-Global-Max (IGM)} principle, which ensures consistency between the optimal joint action and the individual actions of each agent.
This principle is expressed as:
\begin{equation*}
    \arg \max_a Q(s, a) = \left( \arg \max_{a_1} Q_1(\tau_1, a_1), \dots, \arg \max_{a_N} Q_N(\tau_N, a_N) \right),
\end{equation*}
where $Q(s, a)$ is the joint Q-function for all agents, and $Q_i(\tau_i, a_i)$ is the local Q-function for agent $n_i$.
This ensures that during execution, each agent can independently select actions that align with the global objective.
On the other hand, actor-critic methods in CTDE directly optimize agents' policies.
These methods use centralized critics during training, which have access to the global state $s$ or information about other agents, while each agent maintains its own decentralized actor for execution \cite{foerster_CounterfactualMultiAgent_2018, yu_SurprisingEffectiveness_2022, lowe_MultiAgentActorCritic_2017}.

\subsection{Sight Range Dilemma in MARL}\label{sec:bg-sight-range-dilemma}
Recent studies have investigated the impact of sight range on agent coordination within multi-agent reinforcement learning, particularly in relation to communication mechanisms.
For instance, MASIA \cite{guan_EfficientMultiagent_2022} highlights the issue of redundant global information in multi-agent systems, which can hinder effective coordination among agents.
Their experiments in the Traffic Junction \cite{sukhbaatar_LearningMultiagent_2016} environment illustrate that agents utilizing the QMIX algorithm perform better with a limited sight range compared to super-limited or full-sight settings.
CAMA \cite{shao_ComplementaryAttention_2023}, which focuses on dynamic team composition problems, introduces the sight range dilemma and employs attention weights to selectively choose entities to focus on, combined with messages from a global coach.
It demonstrates that both excessive and insufficient information can negatively impact learning.
However, previous works mainly focus on communication environments and utilize communication mechanisms to acquire global information.
In contrast, this paper aims to investigate whether it is possible to address the sight range dilemma directly without the need for additional communication systems.

\subsection{Upper Confidence Bound}\label{sec:bg-ucb}
The \textit{Upper Confidence Bound (UCB)} \cite{auer_FinitetimeAnalysis_2002} algorithm is a widely used method to balance exploration and exploitation in decision-making problems.
At each timestep $t$, UCB selects the action with the highest estimated reward by considering both the empirical mean reward $\hat{X}_t$ and the upper confidence bound of the reward $U_t$, calculated as follows:
\begin{equation}\label{eq:ucb}
    UCB_t(a_i) = \hat{X}_t(a_i) + c \times U_t(a_i) = \hat{X_t}(a_i) + c\times\sqrt\frac{\log t}{N_t(a_i)},
\end{equation}
where $\hat{X}_t(a_i)$ is the empirical mean reward for action $a_i$ up to time $t$, $N_t(a_i)$ is the number of times $a_i$ has been selected, and $c$ is a hyperparameters that controls the exploration and exploitation.
The first term encourages exploitation by favoring actions with higher average rewards, while the second term promotes exploration by choosing actions that have been less frequently explored.

To extend UCB to non-stationary decision-making environments where the optimal choice may change over time, non-stationary UCB algorithms \cite{garivier_UpperConfidenceBound_2008} have been proposed.
Specifically, the \textit{sliding-window upper confidence bound (SW-UCB)} algorithm modifies the calculation of the empirical mean reward in the original UCB by focusing on a limited window of recent observations rather than the entire history:
\begin{equation}\label{eq:sw-ucb}
\begin{aligned}
    UCB^{SW}_t(a_i) &= \hat{X}^{SW}_t(a_i) + c\times U^{SW}_t(a_i), \\
     &= \frac{1}{N_t(a_i, w)}\sum_{j=t-w+1}^t r_j(a_i)\mathbbm{1}_{\{UCB^{SW}_j=i\}} \\
     &\quad+ c\times\sqrt\frac{\log \min(t, w)}{N_t(a_i, w)}\text{, and} \\
    N_t(a_i, w) &= \sum_{j=t-w+1}^t \mathbbm{1}_{\{UCB^{SW}_j=i\}},
\end{aligned}
\end{equation}
where $N_t(a_i, w)$ is the number of times $a_i$ has been selected within the sliding window, $r_j(a_i)$ represents the reward received from selecting action $a_i$ at timestep $j$, and $w$ is a hyperparameter for the size of sliding window.
By limiting the mean reward to the most recent observations, SW-UCB balances adapting to the latest results while maintaining exploration.
This approach makes SW-UCB well-suited for dynamic, non-stationary environments.

\section{Dynamic Sight Range Selection}\label{sec:dsr}
\begin{figure*}
    \centering
    \includegraphics[width=1\linewidth]{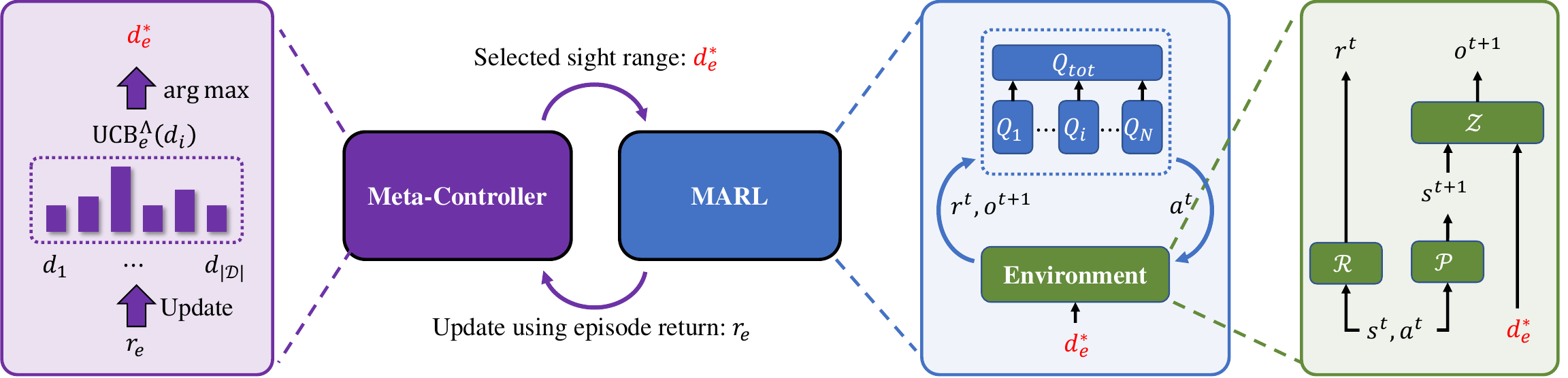}
    \caption{Overview of the Dynamic Sight Range Selection framework. The meta-controller (left) dynamically selects the current optimal sight range $d_e^*$ using the sliding-window UCB based on the episode return $r_e$. The selected sight range $d_e^*$ is used in the MARL training (right), where agents interact with the environment, receiving observations within the selected sight range.}
    \label{fig:alg}
\end{figure*}

In this section, we present our approach, named \textit{Dynamic Sight Range Selection (DSR)}, which dynamically adjusts and finds the most suitable sight range for all agents.
The overview of the DSR method is illustrated in Figure \ref{fig:alg}, which can be incorporated into any MARL algorithm.
In our design, DSR is built upon the Dec-POMDP but modifies the \textit{observation function}, which processes the observation within a given sight range, and introduces a \textit{meta-controller}, which determines the sight range.
Each component is described in detail below.

\subsection{Observation Function}\label{sec:dsr-observation}
The observation function $\mathcal{Z}(s, n_i)$, as reviewed in subsection \ref{sec:bg-marl}, is modified to $\mathcal{Z}(s, n_i, d)$ to incorporate a given sight range $d$, which limits the portion of global state $s$ that each agent $n_i$ can observe.
The sight range $d$ can be defined based on the requirements of each specific environment.
For instance, in Figure \ref{fig:sight_range_dilemma}, the sight range is defined as the distance in grid cells that an agent can observe around its current position, with $d=1$, $d=6$, and $d=3$ for Figure \ref{fig:sight_range_dilemma_1s}, \ref{fig:sight_range_dilemma_6s}, and \ref{fig:sight_range_dilemma_3s}, respectively.
In real-world applications, the sight range may not always be a fixed distance around the agent, and a larger $d$ does not necessarily cover smaller $d$.
For example, in autonomous driving cars, different values of $d$ can represent different sensor designs with varying coverage shapes.
Under this setting, our goal is to find the suitable sight range $d$ during training.

\subsection{Meta-Controller}\label{sec:dsr-meta-controller}

Given a set of sight ranges $\mathcal{D}=\{d_1, d_2, \dots, d_M\}$ with a total of $M$ possible ranges, a meta-controller $\Lambda$ is incorporated into the MARL training by selecting a sight range $d \in \mathcal{D}$ for the agents at the beginning of each episode.
The learning process can be viewed as hierarchical optimization, where the meta-controller dynamically adjusts and selects the optimal sight range at the start of each episode, while the MARL algorithm focuses on maximizing the global value function $Q^{\pi}$ within the selected sight range throughout the episode.
As both the meta-controller and MARL agents learn simultaneously, we adapt the SW-UCB algorithm for the meta-controller to balance exploration and exploitation during training.
The calculation of the UCB score at episode $e$ for each sight range $d_i$ follows equation \eqref{eq:sw-ucb} as:
\begin{equation}\label{eq:meta-controller}
\begin{aligned}
    UCB_e^\Lambda(d_i) &= \hat{X}_e(d_i) + c\times U_e(d_i) \\
            &= \frac{1}{N_e(d_i, w)}\sum_{j=e-w+1}^e r_j(d_i)\mathbbm{1}_{\{UCB_j^\Lambda=i\}}\\
            &\quad+ c\times\sqrt\frac{\log \min(e, w)}{N_e(d_i, w)},
\end{aligned}
\end{equation}
where $e$ represents the $e$-th episode, $w$ is the sliding window size, $r_j(d_i)$ is the episode return with sight range $d_i$ in the $j$-th episode, and $N_e(d_i, w)$ is the number of games that have been played with sight range $d_i$ within the sliding window.
After training, the meta-controller converges to an optimal sight range.
During execution, we simply choose the sight range $d$ with the maximum average return in the window, $\arg\max(\overline{r}(d_i))$, where $\overline{r}(d_i)$ is the average return for each sight range $d_i$.

\subsection{Training Algorithm}\label{sec:dsr-training-algorithm}
Specifically, we summarize the training process in the Algorithm \ref{alg:dsr}.
In line \ref{alg:ucb-selection}, the meta-controller selects the current best sight range $d_e^*$ based on the UCB at the start of each episode.
Once the sight range $d_e^*$ is selected, any MARL algorithm can be applied to train agents for one episode.
The observation is modified based on the selected sight range and used by the agents during interactions with the environment, as shown in line \ref{alg:marl-training-start} to \ref{alg:marl-training-end}.
The modified observation is stored in the buffer associated with the MARL algorithm.
This approach allows the agent to learn from different sight ranges simultaneously, facilitating exploration in the early stages.
As the meta-controller converges, the replay buffer will gradually accumulate more samples with optimized sight ranges.

After training for one episode, the episode return $r_e$ is obtained, and the statistics of $N_e(d_e^*, w)$ and reward for the meta-controller are updated, as shown in line \ref{alg:ucb-update-start} to \ref{alg:ucb-update-end}.
These updated statistics are then used for the next episode.
Overall, the algorithm operates as a hierarchical optimization process, where both the meta-controller and the MARL algorithm are learning simultaneously.
The meta-controller optimizes the sight range selection, while the MARL algorithm focuses on maximizing agent performance within the chosen sight range, leading to efficient coordination and better long-term performance.
Moreover, this design allows the meta-controller to easily integrate with any MARL algorithm.

\begin{algorithm}[h]
\caption{Dynamic Sight Range Selection (DSR)}
\label{alg:dsr}
\begin{algorithmic}[1]
\State \textbf{Input:} Set of sight ranges $\mathcal{D}$, sliding window size $w$, constant $c$, total number of training episodes $E$, total number of training steps per episode $T$
\State \textbf{Output:} The best sight range $d^*$
\For{episode $e = 1$ to $E$}
    \State Meta-controller selects sight range $d_e^*$:\label{alg:ucb-selection}
    \[
    d_e^*= \arg\max_{d_i\in\mathcal{D}} \hat{X}_e(d_i) + c\times U_e(d_i)
    \]
    \State Generate one episode with a modified observation function $\mathcal{Z}(s, n_i, d_e^*)$\label{alg:marl-training-start}
    \State Obtain the episode return $r_e$
    \State Train the episode by any MARL algorithm \label{alg:marl-training-end}
    \If{$d_{e-t} \neq d_e^*$}\label{alg:ucb-update-start}
        \[
        N_e(d_e^*, w) = N_e(d_e^*, w) + 1
        \]
    \EndIf    
    \State Record reward $r_e$ for $d_e^*$\label{alg:ucb-update-end}
\EndFor
\end{algorithmic}
\end{algorithm}

\section{Experiment}\label{sec:exp}

\subsection{Experiment Setup}\label{sec:exp-setup}
\begin{figure*}[h]
  \centering
  \begin{minipage}{0.24\textwidth}
    \centering
    \includegraphics[width=\linewidth,height=4cm,keepaspectratio]{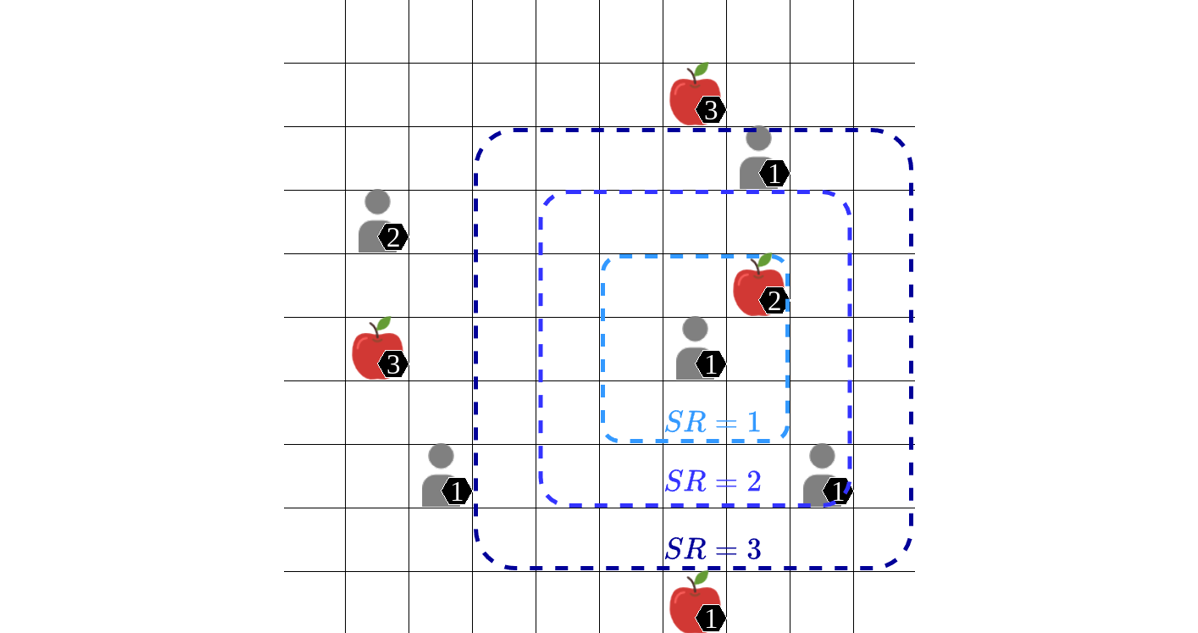}
    \subcaption{Level-Based Foraging (LBF)}
    \label{fig:lbf-env-sr}
  \end{minipage}
  \hfill
  \begin{minipage}{0.24\textwidth}
    \centering
    \includegraphics[width=\linewidth,height=4cm,keepaspectratio]{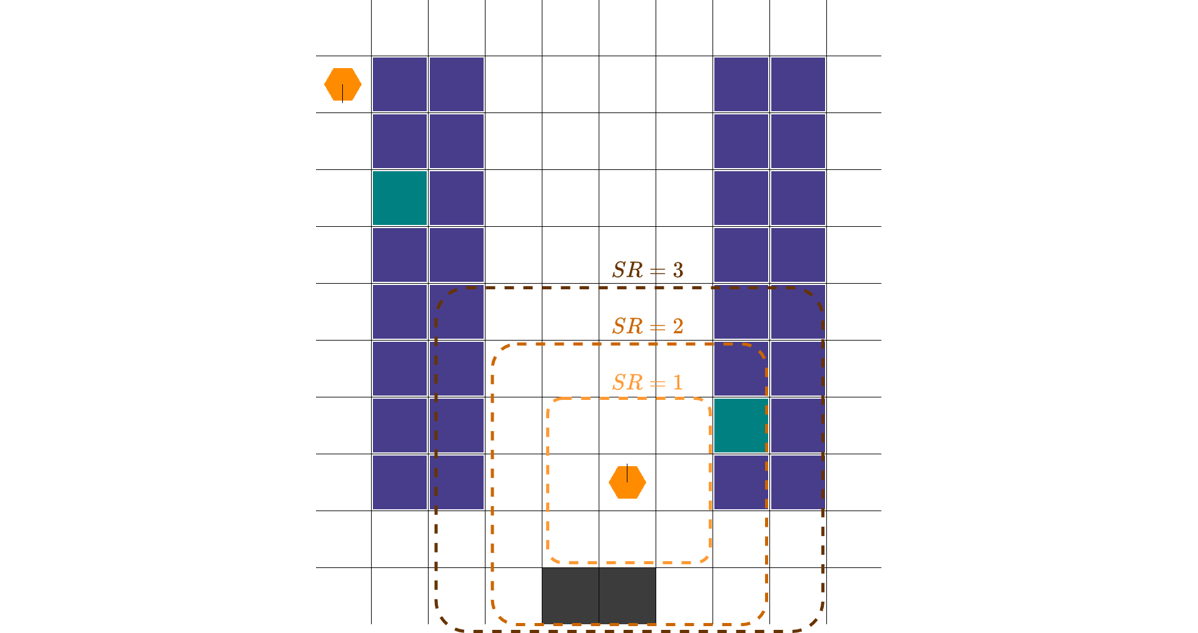}
    \subcaption{Multi-Robot Warehouse (RWARE)}
    \label{fig:rware-env-sr}
  \end{minipage}
  \hfill
  \begin{minipage}{0.48\textwidth}
    \centering
    \includegraphics[width=0.48\linewidth,height=4cm,keepaspectratio]{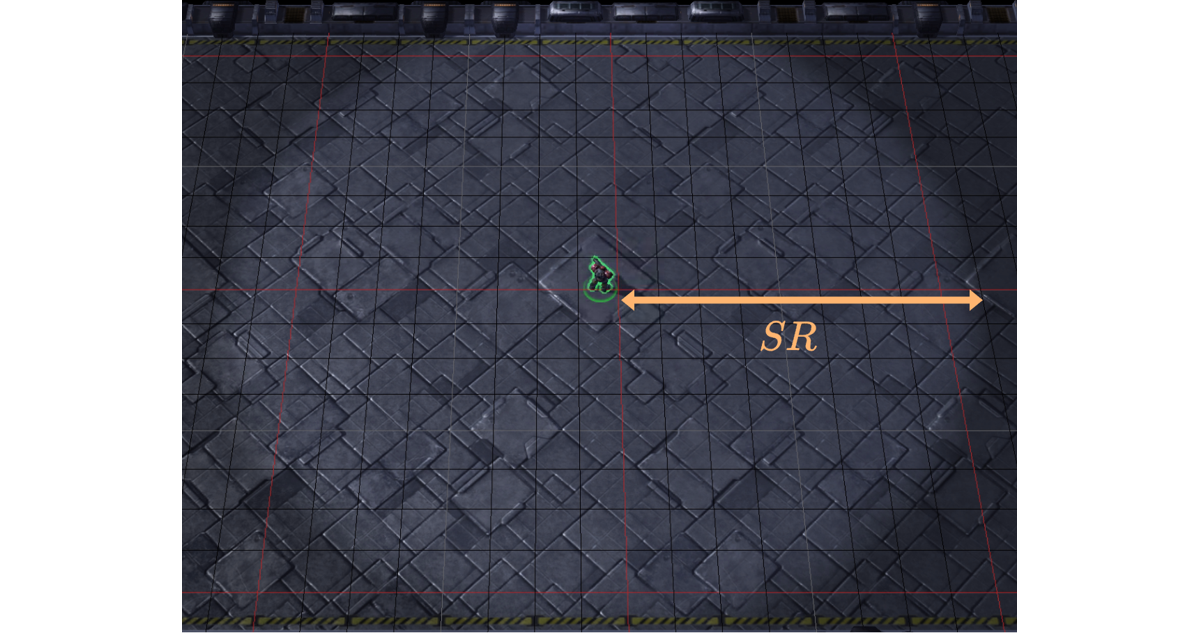}
\includegraphics[width=0.48\linewidth,height=4cm,keepaspectratio]{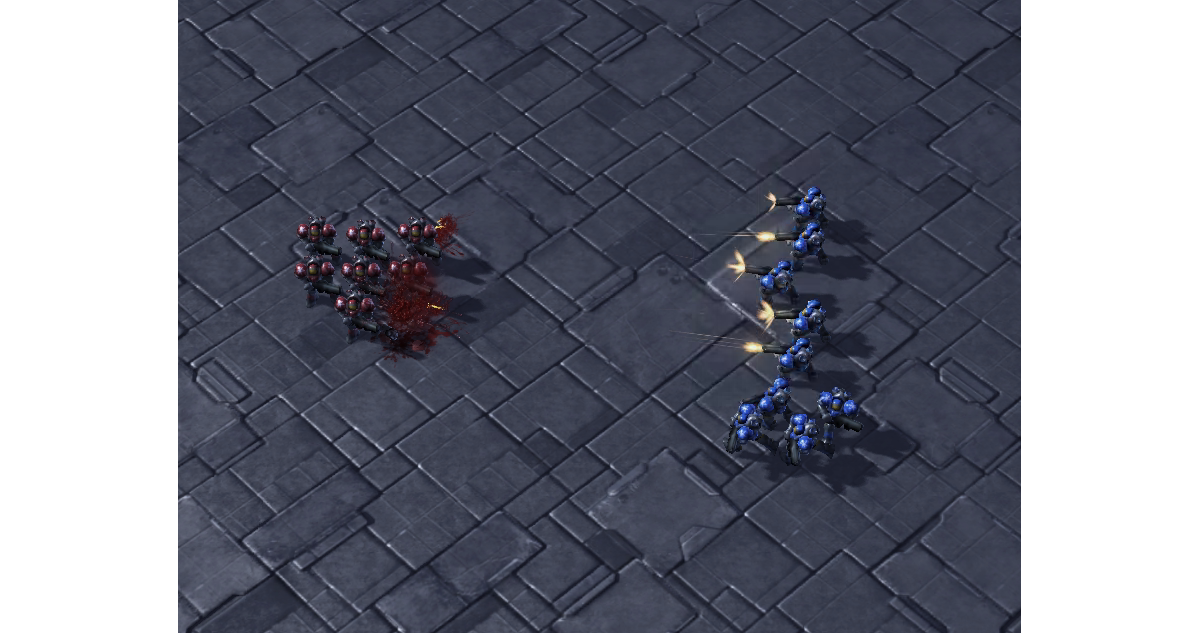}
    \subcaption{StarCraft Multi-Agent Challenge (SMAC). Left: An Example of sight range; right: a game example (8m\_vs\_9m)}
    \label{fig:smac-env}
  \end{minipage}
  \caption{MARL environments used in our experiments.}
  \label{fig:each_env}
\end{figure*}

We build DSR upon the training framework EPyMARL \cite{papoudakis_BenchmarkingMultiAgent_2021} and conduct experiments in three common MARL environments, as illustrated in Figure \ref{fig:each_env}, including Level-Based Foraging (LBF) \cite{albrecht_GametheoreticModel_2013, papoudakis_BenchmarkingMultiAgent_2021}, Multi-Robot Warehouse (RWARE) \cite{papoudakis_BenchmarkingMultiAgent_2021}, and the StarCraft Multi-Agent Challenge (SMAC) \cite{samvelyan_StarCraftMultiAgent_2019}.
For the SW-UCB in the meta-controller, we choose $c=2$ for the exploration coefficient and $w=5000$ for the sliding window size.
The following paragraphs describe each environment and its specific settings.

\paragraph{\textbf{Level-Based Foraging (LBF)}}
LBF is a grid-based multi-agent environment designed to evaluate cooperative behavior, where agents must collaborate to collect food in a grid world, as shown in Figure \ref{fig:lbf-env-sr}, making it a common benchmark for testing MARL algorithms.
The episode return ranges from 0 to 1 and is proportional to the fraction of food collected relative to the total food score on the map.
The sight range is defined by the distance (in grid cells) that an agent can observe around its current position.
For example, a sight range of 1 provides a 3x3 ($1+1\times2=3$) view around the agent, while a sight range of 2 expands this to a 5x5 ($1+2\times2=5$) area.
In LBF, the notation 10x10-4p-2f-coop represents a 10x10 grid map, four players (4p), two food items (2f), and a cooperative mode (coop), where agents must cooperate more intensively to collect the food.
For our experiments, we choose three settings: 10x10-4p-2f-coop, 10x10-4p-2f, and 10x10-4p-4f-coop.
For each setting, we select two different default sight ranges, $d=6$ and $d=10$.
Note that in a 10x10 grid map, a sight range of $d=10$ allows each agent to observe the entire map, equivalent to having access to the global state.
For the DSR method, the meta-controller can select the sight range from the set $\mathcal{D}$ as follows:
\begin{itemize}
    \item $\mathcal{D}=\{2,4,6\}$ for $d=6$
    \item $\mathcal{D}=\{2,4,6,8,10\}$ for $d=10$
\end{itemize}

\paragraph{\textbf{Multi-Robot Warehouse (RWARE)}}
RWARE is a grid-based multi-agent environment where cooperative robots work together to transport goods within a warehouse.
Agents must locate and deliver requested shelves (girds marked in teal) to the designated locations (grids marked in black at the bottom), and then return the shelves to empty positions before continuing with the next delivery, as shown in Figure \ref{fig:rware-env-sr}.
Each successful delivery yields a reward of +1, incentivizing agents to complete delivery cycles.
Since both RWARE and LBF are grid-based environments, the sight range in RWARE is defined in the same way as LBF.
In RWARE, we use two map sizes, a 10x11 grid map (tiny) and a 10x20 (small) map, each with two agents (denoted as tiny-2ag and small-2ag).
For each setting, we select two different default sight ranges, $d=3$ and $d=5$.
For the DSR method, the meta-controller can select the sight range from the set $\mathcal{D}$ as follows:
\begin{itemize}
    \item $\mathcal{D}=\{1,2,3\}$ for $d=3$
    \item $\mathcal{D}=\{1,2,3,4,5\}$ for $d=5$
\end{itemize}

\paragraph{\textbf{StarCraft Multi-Agent Challenge (SMAC)}}
SMAC is a well-established benchmark for multi-agent reinforcement learning, focusing on micromanagement tasks in StarCraft II. 
Each agent controls a unique unit, and the goal is to defeat AI-controlled opponents in combat, as shown in Figure \ref{fig:smac-env}.
The sight range is defined as the visibility radius around each unit.
We select six settings for SMAC, including 5m\_vs\_6m, 8m\_vs\_9m, 10m\_vs\_11m, 3s\_vs\_5z, 3s5z, and MMM2.
For each setting, we select three different default sight ranges, $d=9$, $d=15$, and $d=21$.
For the DSR method, the meta-controller can select the sight range from the set $\mathcal{D}$ as follows:
\begin{itemize}
    \item $\mathcal{D}=\{3,6,9\}$ for $d=9$
    \item $\mathcal{D}=\{3,6,9,12,15\}$ for $d=15$
    \item $\mathcal{D}=\{3,6,9,12,15,18,21\}$ for $d=21$
\end{itemize}

In LBF and RWARE, the global state $s$ provided by the environment does not contain all information.
This mirrors real-world applications, where obtaining complete global information is often impractical \cite{hu_MARLlibScalable_2023, terry_PettingzooGym_2021}.
A common approach is to approximate the global state by concatenating the observations from all agents.
Although the original SMAC environment provides complete global information, our focus is on how sight ranges influence the learning complexity of observations and states.
Therefore, we adopt a variant of SMAC where the global state is derived by concatenating the observations from all agents, the same approach used in LBF and RWARE.

\subsection{Performance of DSR}\label{sec:exp-dsr-performance}
We first train QMIX \cite{rashid_QMIXMonotonic_2018}, a common MARL algorithm, with and without DSR across the three environments with various settings described in the previous subsection.
Each setting was trained with five different seeds.

Table \ref{tab:overall-performance} shows the results comparing DSR and baseline (without DSR) across different environment settings.
For LBF and RWARE, we use the mean test return, while for SMAC, we use the mean test win rate.
Note that the last notation after the hyphen (e.g., the ``10s" in 10x10-4p-2f-coop-10s) represents the default sight range used in the baseline (without DSR).
For DSR, the set of sight ranges is as mentioned in the previous subsection.

\begin{figure}[h]
    \centering
    \includegraphics[width=\linewidth]{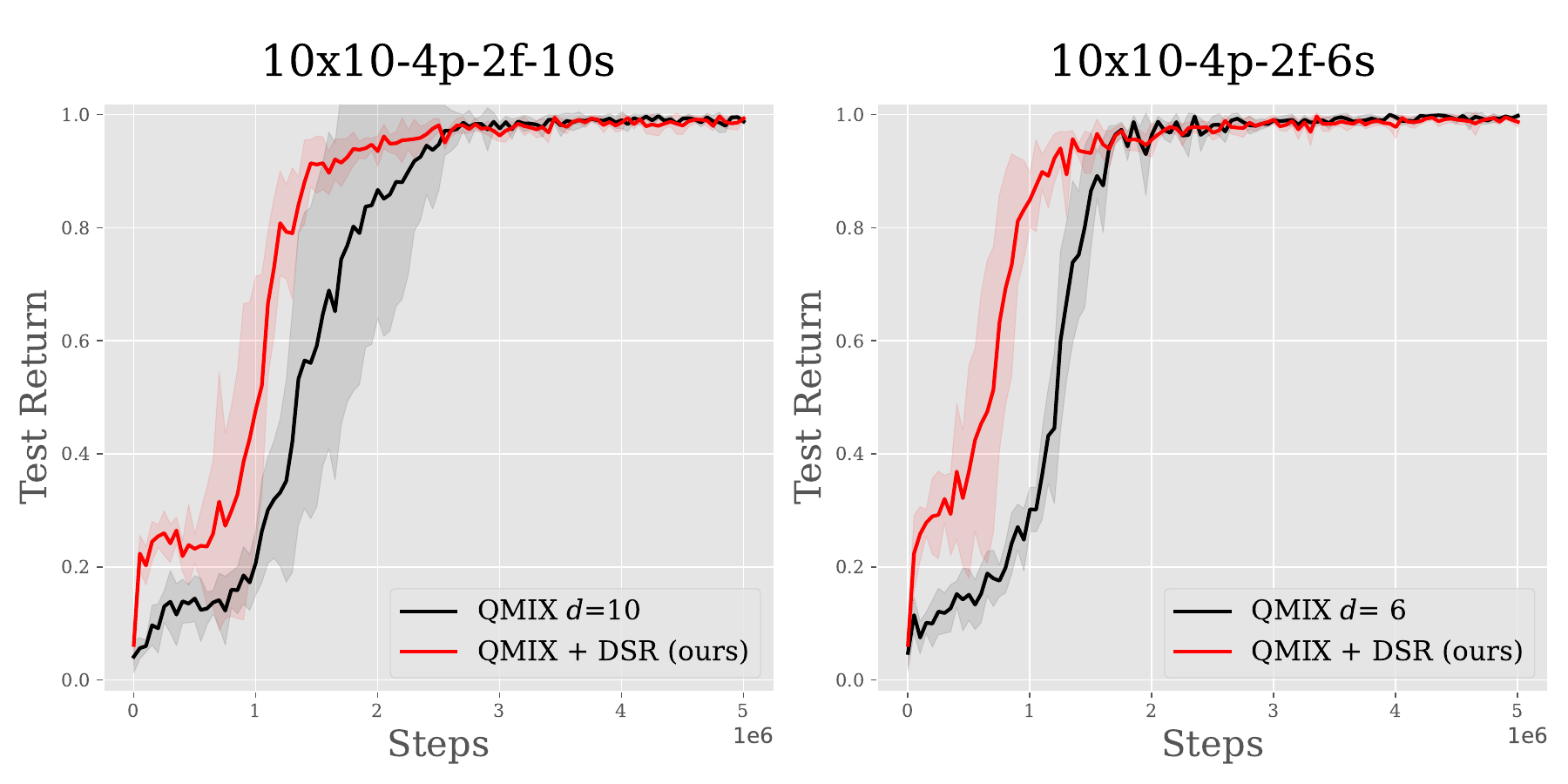}
    \caption{Mean test returns on two LBF environment settings. The shaded area represents the standard deviation.}
    \label{fig:lbf-dsrVsBaseline}
\end{figure}

Our results show that DSR consistently improves performance across all three environments.
In LBF, our method significantly outperforms the baseline, with substantial improvement in complex cooperative settings such as 10x10-4p-4f-coop-6s (0.772 compared to 0.277) and 10x10-4p-4f-coop-10s (0.798 compared to 0.338).
This result demonstrates the sight range dilemma, where larger sight ranges in complex environments are not always beneficial, as agents may receive excessive irrelevant information that negatively impacts decision-making, as illustrated in Figure \ref{fig:sight_range_dilemma}.
The scores for LBF range from 0 to 1, so in some simple settings both w/ and w/o DSR can achieve nearly optimal scores. However, even in these cases, DSR significantly accelerates the training process, as shown in Figure \ref{fig:lbf-dsrVsBaseline}. For more challenging LBF tasks, such as 10x10-4p-4f-coop-10s, which require deeper cooperation, DSR demonstrates substantial improvements (0.798 vs. 0.338) compared to training without DSR.
For RWARE, both models struggle in the small map, likely due to the map's difficulty, achieving relatively lower scores, but DSR outperforms the baseline on the tiny map.
In SMAC, DSR consistently enhances performance, particularly in scenarios with larger sight range settings, such as 3s\_vs\_5z-21s (0.712 compared to 0.292) and MMM2-21s (0.714 compared to 0.19).
In conclusion, these results demonstrate the effectiveness of dynamically selecting sight ranges using DSR, highlighting its robustness across a wide range of environments.

\begin{table}[]\centering
\caption{Comparison between baseline and DSR across three environments.}\label{tab: }
\begin{small}
\begin{tabular}{llrr}\toprule
& &w/o DSR & w/ DSR (ours)\\\midrule
\multirow{6}{*}{LBF} &10x10-4p-2f-coop-10s &0.769 ± 0.384 &\textbf{0.925 ± 0.064} \\
&10x10-4p-2f-coop-6s &\textbf{0.972 ± 0.037} &0.957 ± 0.040 \\
&10x10-4p-2f-10s &0.988 ± 0.005 &\textbf{0.993 ± 0.004} \\
&10x10-4p-2f-6s &\textbf{0.998 ± 0.004} &0.987 ± 0.010 \\
&10x10-4p-4f-coop-10s &0.338 ± 0.339 &\textbf{0.798 ± 0.050} \\
&10x10-4p-4f-coop-6s &0.277 ± 0.194 &\textbf{0.772 ± 0.068} \\
\midrule
\multirow{4}{*}{RWARE} &tiny-2ag-5s &1.486 ± 1.361 &\textbf{4.762 ± 4.702} \\
&tiny-2ag-3s &5.846 ± 1.426 &\textbf{11.900 ± 6.474} \\
&small-2ag-5s &\textbf{0.074 ± 0.148} &0.050 ± 0.100 \\
&small-2ag-3s &\textbf{0.182 ± 0.359} &0.036 ± 0.072 \\
\midrule
\multirow{18}{*}{SMAC} 

&5m\_vs\_6m-9s & 0.102 ± 0.028 & \textbf{0.128 ± 0.066} \\
&5m\_vs\_6m-15s &0.080 ± 0.100 & \textbf{0.140 ± 0.055} \\
&5m\_vs\_6m-21s &0.054 ± 0.079 & \textbf{0.112 ± 0.053} \\

&8m\_vs\_9m-9s & 0.452 ± 0.114 & \textbf{0.508 ± 0.077} \\
&8m\_vs\_9m-15s & 0.234 ± 0.287 & \textbf{0.504 ± 0.087} \\
&8m\_vs\_9m-21s & 0.120 ± 0.190 & \textbf{0.472 ± 0.078} \\

&10m\_vs\_11m-9s & 0.402 ± 0.220 & \textbf{0.540 ± 0.061} \\
&10m\_vs\_11m-15s & 0.122 ± 0.202 & \textbf{0.546 ± 0.127} \\
&10m\_vs\_11m-21s & 0.186 ± 0.262 & \textbf{0.680 ± 0.066} \\

&3s\_vs\_5z-9s & \textbf{0.716 ± 0.143} & 0.676 ± 0.065  \\
&3s\_vs\_5z-15s & 0.498 ± 0.409 & \textbf{0.660 ± 0.081} \\
&3s\_vs\_5z-21s & 0.292 ± 0.260 & \textbf{0.712 ± 0.110} \\

&3s5z-9s &\textbf{ 0.854 ± 0.086} & \textbf{0.854 ± 0.051} \\
&3s5z-15s &\textbf{0.808 ± 0.094} & 0.770 ± 0.075 \\
&3s5z-21s & \textbf{0.784 ± 0.156} & 0.736 ± 0.069 \\

&MMM2-9s & 0.690 ± 0.122 & \textbf{0.736 ± 0.076} \\
&MMM2-15s & 0.572 ± 0.113 & \textbf{0.648 ± 0.094} \\
&MMM2-21s & 0.190 ± 0.266 & \textbf{0.714 ± 0.101} \\

\bottomrule
\end{tabular}
\end{small}
\label{tab:overall-performance}
\end{table}

In addition, we found that using DSR can accelerate the training process.
For example, Figure \ref{fig:lbf-dsrVsBaseline} shows the training curves for two LBF settings, including 10x10-4p-2f-10s and 10x10-4p-2f-6s.
While both DSR and the baseline converge to similar results by the end of the training, the training curve of DSR (red) rises more quickly during the early training steps and remains higher or equal to the baseline (black) throughout the training process.
This demonstrates an additional benefit of using DSR, as training with different sight ranges helps agents learn the game more quickly.

\subsection{The Sight Range Dilemma}\label{sec:exp-sight-range-dilemma}
\begin{figure}[!ht] 
    \centering
    \begin{subfigure}[]{0.45\textwidth}
        \centering
        \includegraphics[width=0.48\textwidth]{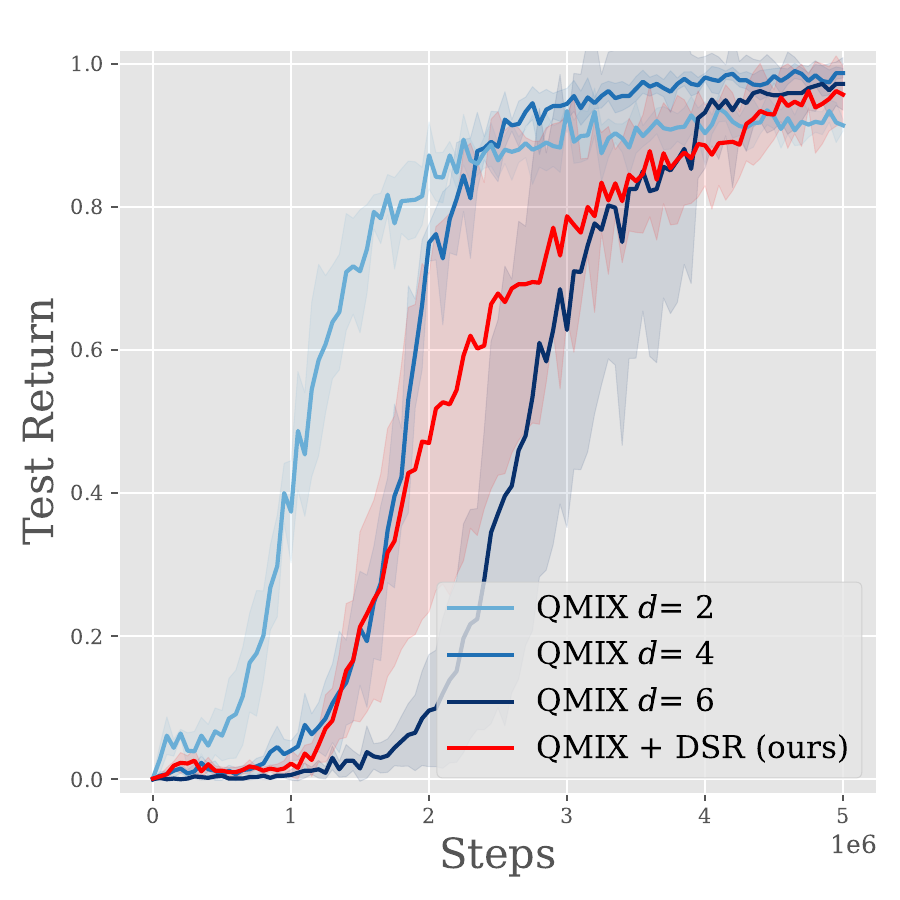}
        \includegraphics[width=0.48\textwidth]{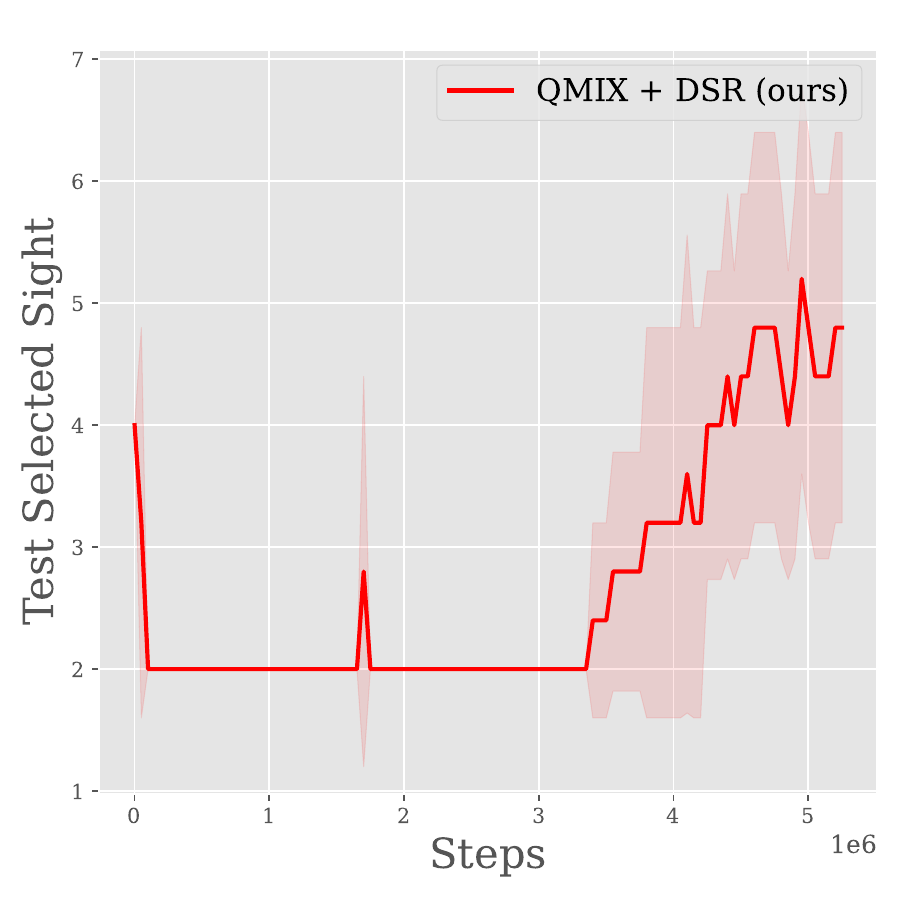}
        \caption{LBF 10x10-4p-2f-coop-6s}
        \label{fig:lbf-DsrAndPuresAndSelected}
    \end{subfigure}
    \\
    \begin{subfigure}[]{0.45\textwidth}
        \centering
        \includegraphics[width=0.48\textwidth]{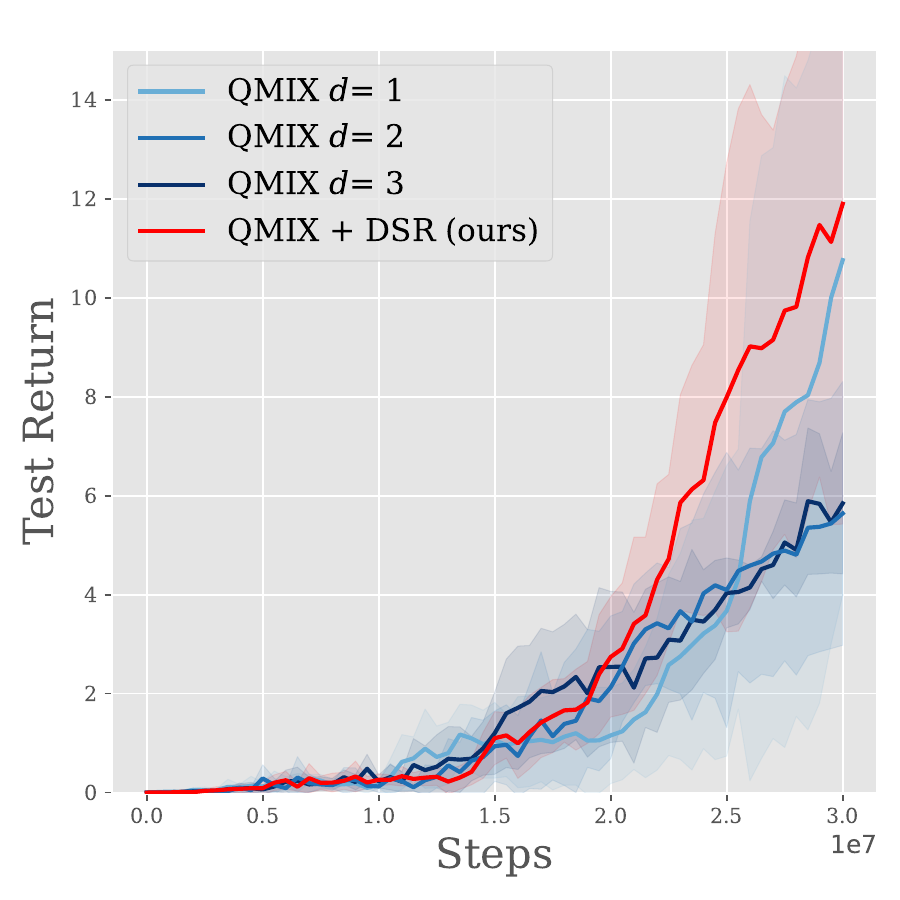}
        \includegraphics[width=0.48\textwidth]{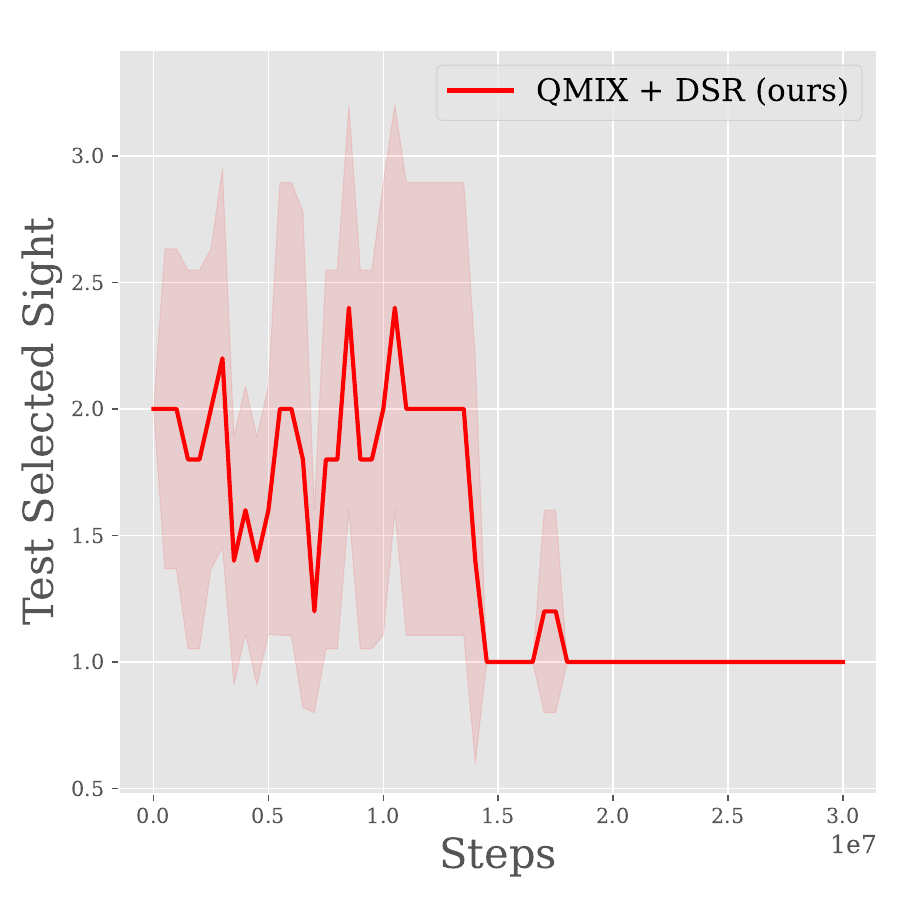}
        \caption{RWARE tiny-2ag-3s}
        \label{fig:rware-DsrAndPuresAndSelected}
    \end{subfigure}
    \\
    \begin{subfigure}[]{0.45\textwidth}
        \centering
        \includegraphics[width=0.48\textwidth]{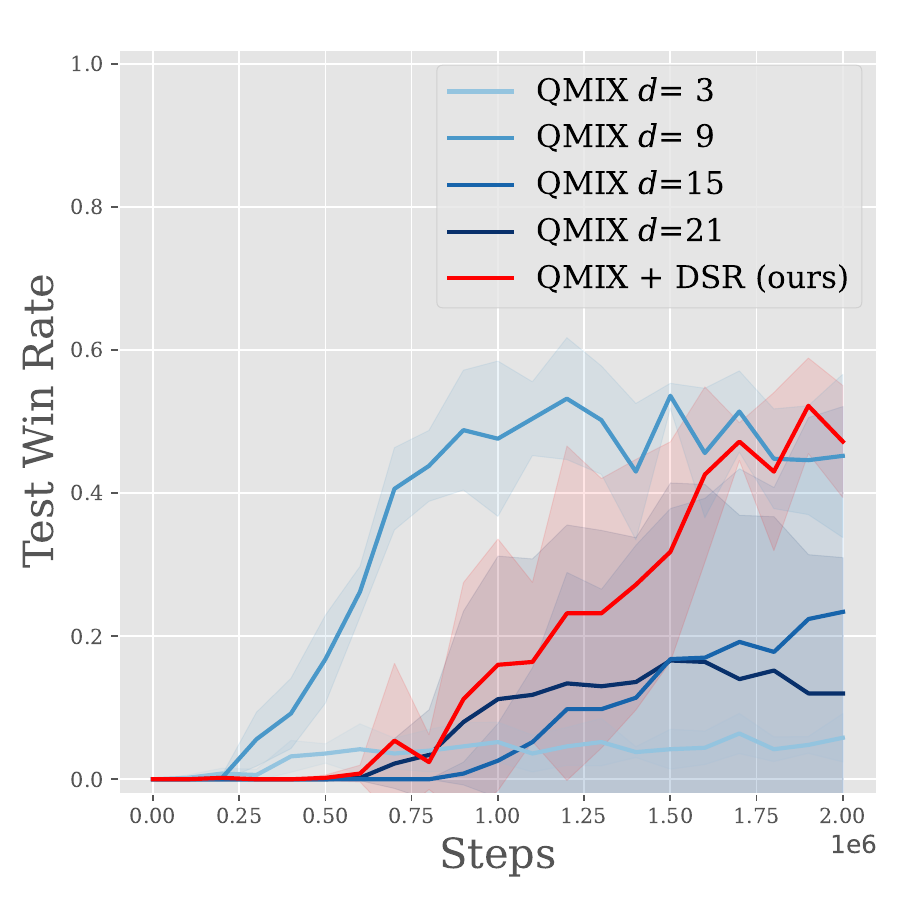}
        \includegraphics[width=0.48\textwidth]{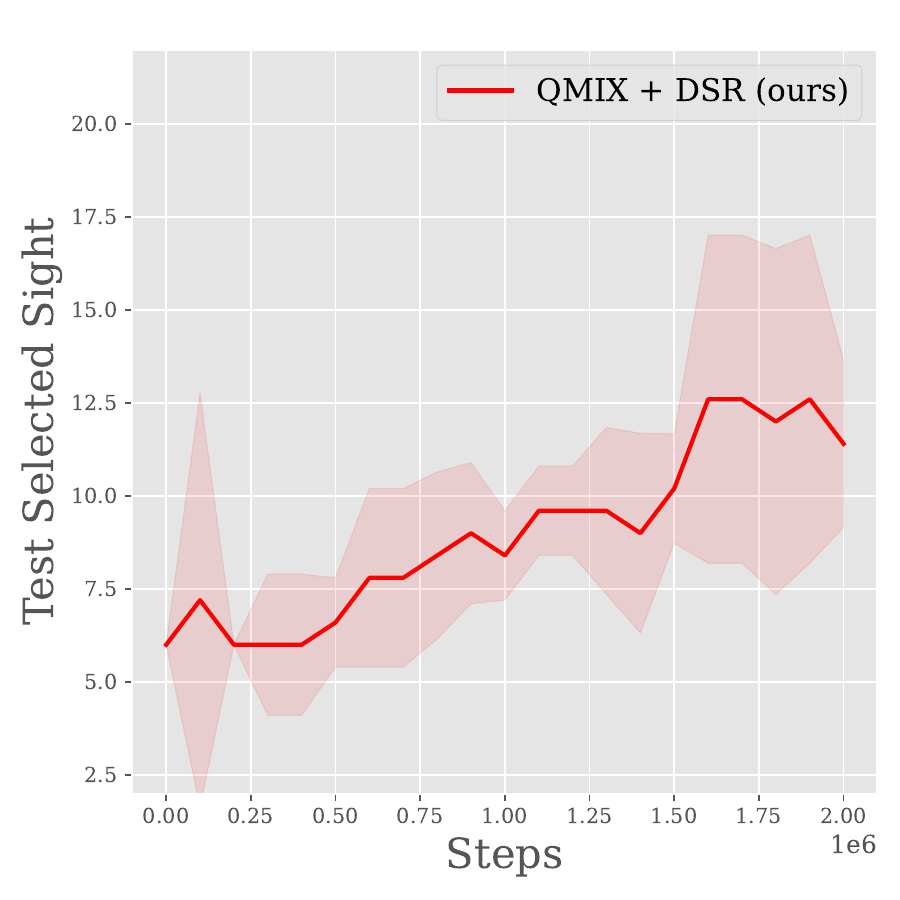}
        \caption{SMAC 8m\_vs\_9m-21s}
        \label{fig:smac-DsrAndPuresAndSelected}
    \end{subfigure}
    \caption{Experiments on three environment settings. For each subfigure, the left shows the mean test returns, while the right side displays the selected sights by DSR during training.}
    \label{fig:combined-learning-curves}
\end{figure}

To further investigate how sight range affects training, we conduct an additional experiment where the baseline uses different fixed sight ranges throughout the entire training process.
We select one setting for each environment, as illustrated in Figure \ref{fig:combined-learning-curves}.
Figure \ref{fig:lbf-DsrAndPuresAndSelected} shows the results for the LBF environment.
The left subfigure shows the mean test return curves of the baseline with different fixed sight ranges ($d=2, 4,\text{ and }6$) and our DSR approach, while the right subfigure shows the sight range dynamically selected by DSR during the training process.
From the left subfigure, we observe that using a smaller sight range results in higher returns at the early training.
For instance, at training step $2\times10^6$, the performance of $d=2$ is better than $d=4$, and $d=4$ outperforms $d=6$.
However, by the end of the training, $d=2$ plateaus, while $d=4$ and $d=6$ achieve better performance.
Interestingly, since our DSR can dynamically adjust the sight range, we observe that DSR tends to select smaller sight ranges at the beginning and gradually shifts to larger sight ranges, as shown in the right subfigure.
The RWARE environment, as shown in Figure \ref{fig:rware-DsrAndPuresAndSelected}, shows another interesting result.
In the left subfigure, we observe that using $d=1$ performs well in this environment setting.
Surprisingly, in the right subfigure, DSR initially attempts to explore a larger sight range but then quickly converges to $d=1$.
In the SMAC environment, as shown in figure \ref{fig:smac-DsrAndPuresAndSelected}, we observe that DSR gradually selects sight range from $d=6$ to $d=12$, and the results outperform all fixed sight ranges.
This suggests that agents benefit from training with smaller sight ranges initially, which helps them transition more effectively to larger sight ranges during training.

\begin{figure}
    \centering
    \includegraphics[width=0.48\linewidth]{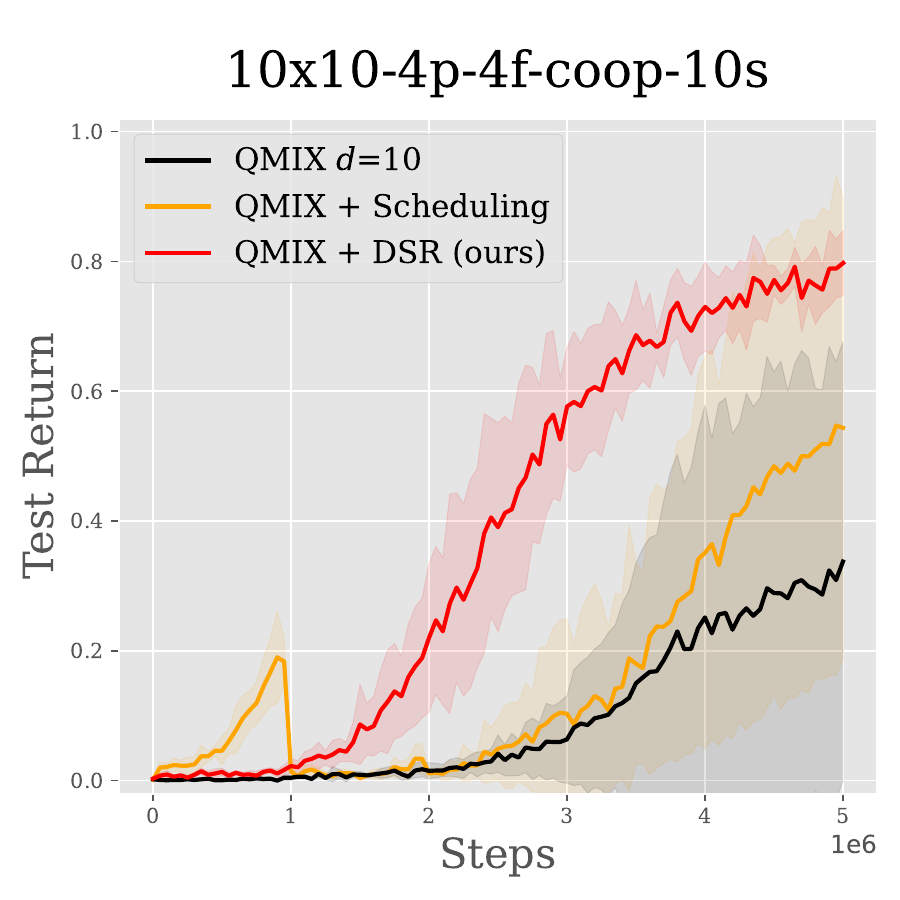}
    \caption{Comparison between the baseline, DSR, and a fixed sight range scheduling approach in the LBF environment.}
    \label{fig:lbf-POEM}
\end{figure}

Moreover, in LBF, we observe that DSR usually starts by selecting smaller sight ranges and then gradually selects larger ones.
A straightforward approach would be to design a fixed sight range scheduling, such as the sight range progressively expanding from small to large (e.g., $d=2, 4, 6, 8,\text{ and }10$) at regular intervals, dividing the training steps into five equal phases.
As shown in Figure \ref{fig:lbf-POEM}, the fixed sight range scheduling approach also performs better than the baseline but still does not outperform DSR.
Additionally, when switching sight ranges in the fixed schedule, we observe a noticeable drop in performance.
In contrast, DSR allows for smoother sight range adjustments and results in a more stable training curve.

In summary, our results demonstrate that DSR can not only accelerate the training but also automatically discover the appropriate sight range without the need to manually train across all sight ranges.
In many real-world complex environments, different sight ranges may yield varying outcomes and it is often difficult to know the best sight range for every task.
Therefore, DSR offers an efficient solution for both finding the optimal sight range and subtly addressing the sight range dilemma.

\subsection{DSR in Other MARL Algorithms}\label{sec:exp-dsr-generalization}
Since DSR does not modify the underlying MARL algorithms, we conduct experiments to verify its generalizability across different MARL algorithms, including IQL, VDN, IPPO, and MAPPO, in LBF and RWARE environments.
In general, the sight range dilemma persists across all algorithms, and DSR consistently outperforms the baseline, similar to the results shown for QMIX (subsection \ref{sec:exp-dsr-performance}).

\begin{figure}[htb]
    \centering
    \begin{subfigure}{0.5\columnwidth}
        \centering
        \includegraphics[height=0.6\textheight, keepaspectratio]{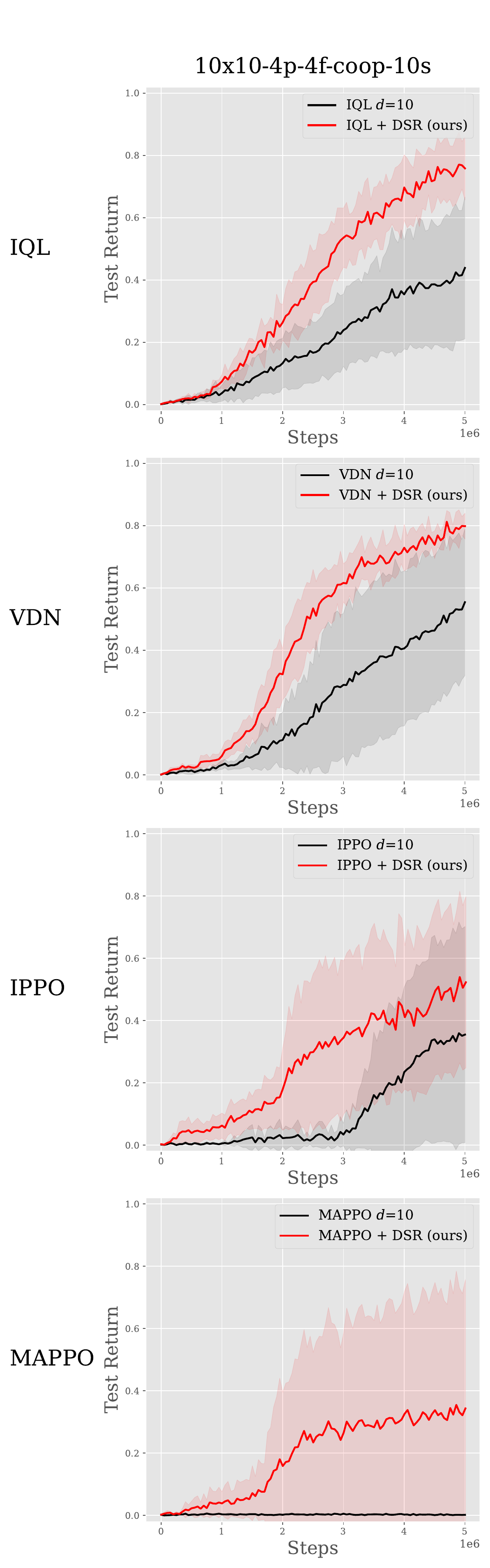}
    \end{subfigure}%
    \begin{subfigure}{0.5\columnwidth}
        \centering
        \includegraphics[height=0.6\textheight, keepaspectratio]{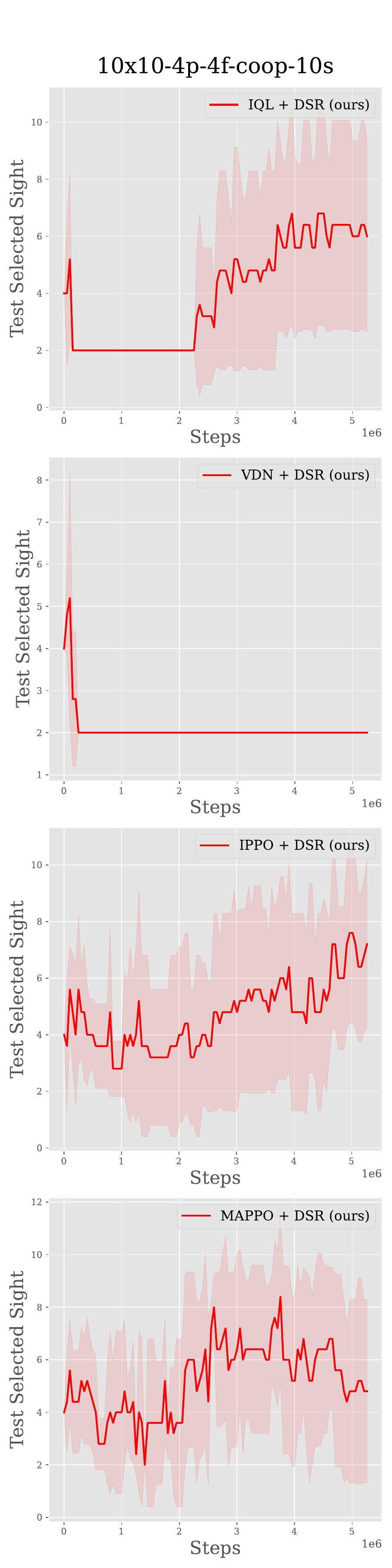}
    \end{subfigure}
    \caption{Experiments with four additional MARL algorithms. Top: Mean test returns for the baseline and DSR; Bottom: Sight ranges selected by DSR during training.}
    \label{fig:dsr-moreAlg}
\end{figure}

Figure \ref{fig:dsr-moreAlg} shows one of the results for the LBF environment with the 10x10-4p-4f-coop setting, where DSR significantly outperforms the baseline across all four algorithms.
In addition, we observe that DSR follows a similar pattern of adjusting sight ranges across most algorithms, except for VDN, typically starting with smaller sight ranges and gradually transitioning to larger ones.
These experiments further demonstrate the robustness and versatility of DSR, showing that our approach can be easily integrated into any MARL algorithm and enhance learning performance without requiring algorithm-specific modifications.

\subsection{Exploring Hyperparameters in DSR}\label{subsec:ucb-ablation}
\begin{figure}[htb]
  \centering
  \begin{subcaptionblock}{0.32\columnwidth}
    \includegraphics[width=\textwidth]{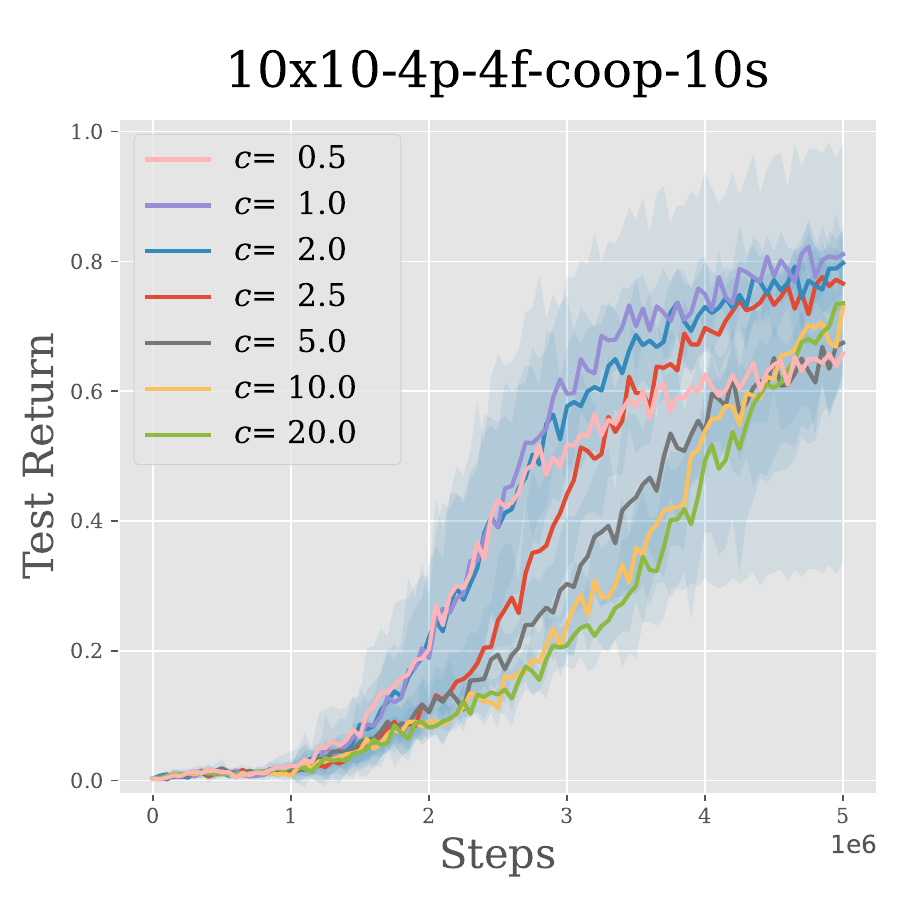}
    \caption{Exploration coefficient $c$.}
    \label{fig:lbf-ucb-ablation-each_env-c}
  \end{subcaptionblock}
  \begin{subcaptionblock}{0.32\columnwidth}
    \includegraphics[width=\textwidth]{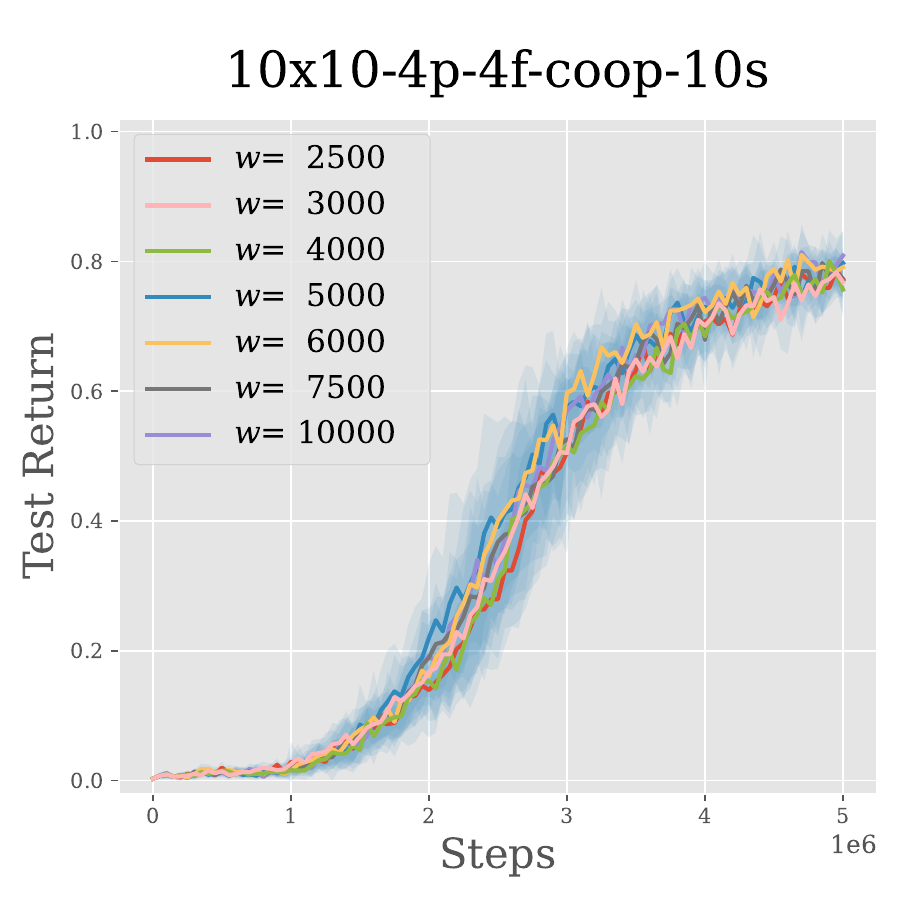}
    \caption{Sliding window size $w$.}
    \label{fig:lbf-ucb-ablation-each_env-w}
  \end{subcaptionblock}
  \begin{subcaptionblock}{0.32\columnwidth}
    \includegraphics[width=\textwidth]{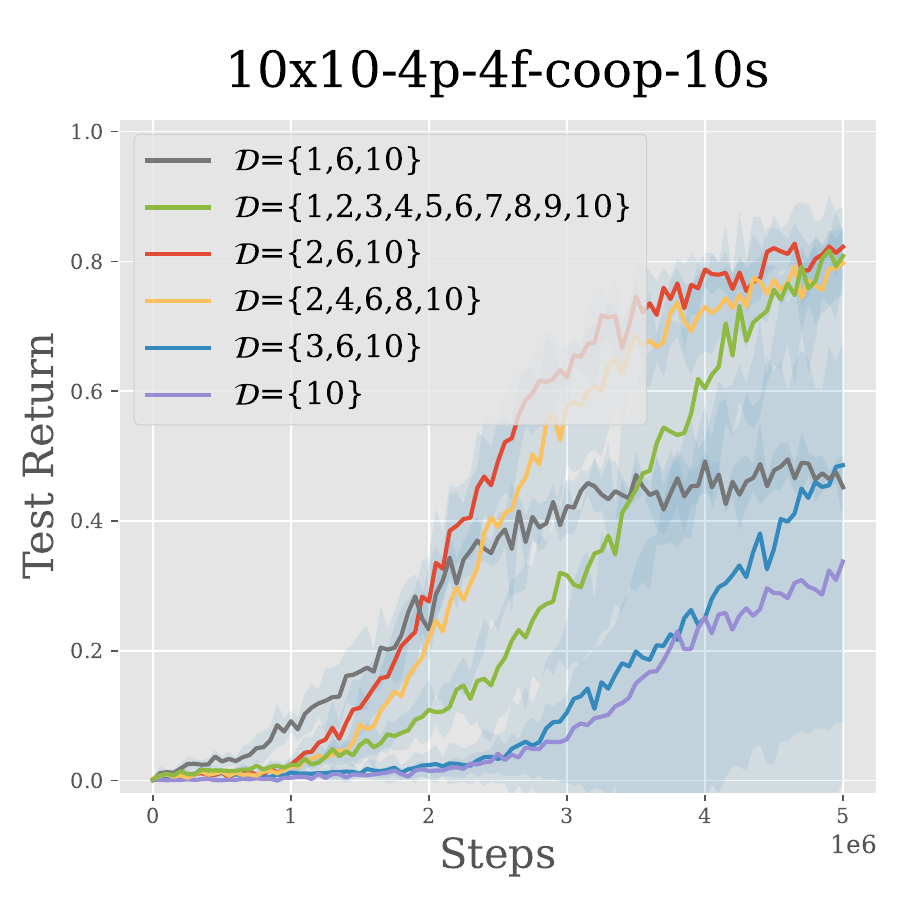}
    \caption{Different set of sight range $\mathcal{D}$.}
    \label{fig:lbf-ucb-ablation-each_env-sightList}
  \end{subcaptionblock}
  \caption{Experiments for different hyperparameters in the LBF environment.}
  \label{fig:lbf-ucb-ablation-each_env}
\end{figure}

In this subsection, we analyze the effectiveness of different hyperparameters in DSR within the LBF environment, including the exploration constant $c$, the sliding window size $w$, and various combinations of sight ranges available for the meta-controller to select.
The results are shown in Figure \ref{fig:lbf-ucb-ablation-each_env}.
For the exploration coefficient $c$, we observe that neither smaller nor larger values consistently lead to better results, as shown in Figure \ref{fig:lbf-ucb-ablation-each_env-c}.
The findings suggest that the optimal range for $c$ lies between 1 and 2.5.
In contrast, the choice of sliding window size does not significantly affect the results, as shown in \ref{fig:lbf-ucb-ablation-each_env-w}.

Next, we analyze the performance when different sets of sight ranges are used in DSR, as shown in Figure \ref{fig:lbf-ucb-ablation-each_env-sightList}.
When comparing $\mathcal{D}=\{1,2,3,\dots,10\}$ (green) and $\mathcal{D}=\{2,4,6,8,10\}$ (yellow), the results show that a larger set of options for meta-controller slows down the training process, as it increases the complexity of sight range selection for meta-controller and requires more time for agents to adapt.
For $\mathcal{D}=\{2,6,10\}$ (red), $\mathcal{D}=\{1,6,10\}$ (gray), and $\mathcal{D}=\{3,6,10\}$ (blue), the results show that the red line outperforms the other two.
This is because $d=2$ enables faster learning during the early stage in LBF, as illustrated in Figure \ref{fig:lbf-DsrAndPuresAndSelected}.
However, in practice, it is difficult to predict which sight range is best in advance.
In summary, a larger set of sight ranges is less likely to miss potentially effective sight ranges, but it may increase training time due to exploration. 
On the other hand, a smaller sight range can learn faster if it includes an appropriate sight range, but it also risks omitting better ones. 
Therefore, designing an appropriate set of sight range $\mathcal{D}$ is important.

\subsection{Comparison to Communication-based Methods}
Finally, we compare our DSR approach to CAMA, a communication-based method that uses an attention-weight ranking mechanism to select relevant entities, aiming to address the sight range dilemma.
To evaluate DSR, we integrate it into the source code provided in CAMA \cite{shao_ComplementaryAttention_2023}.
We then follow CAMA's training process using the IM-Qatten algorithm, which combines Qatten \cite{yang_QattenGeneral_2020} with an inverse model \cite{pathak_CuriositydrivenExploration_2017}.
In addition, we use the same environment as CAMA, SMAC Dynamic Team Composition (SMAC-DT), a variant of SMAC where both the number and types of units change dynamically across episodes.
The number of agents is randomly set between 3 and 5 during both training and testing.
The default sight range in CAMA is set to $d=9$.
For the DSR method, the meta-controller can select the sight range from the set $\mathcal{D}=\{3,6,9\}$.

\begin{figure}[htbp] 
    \centering
    \includegraphics[width=\linewidth]{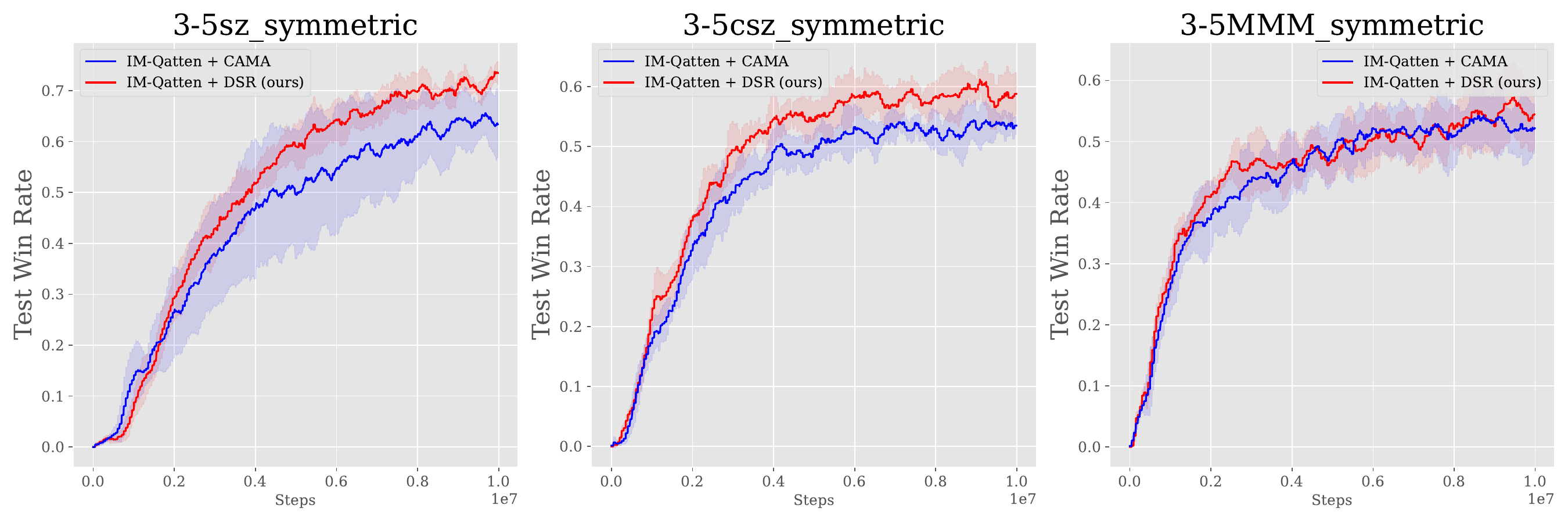}  
    \caption{Comparison between CAMA and DSR in the SMAC-DT environment.}
    \label{fig:smac-dt-win}  
\end{figure}

Figure \ref{fig:smac-dt-win} shows that DSR outperforms the attention-weight ranking mechanism across three SMAC-DT environment settings.
DSR also provides explanability, as it explicitly indicates which sight range is selected, while CAMA's attention-weight ranking can only show the chosen entities without offering insight into the overall observation strategy.
Overall, this experiment demonstrates the advantages of DSR in both performance and interpretability.

\section{Discussion}
This paper proposes a novel approach named Dynamic Sight Range Selection (DSR), which successfully addresses the sight range dilemma issues.
Our experiments show that DSR outperforms the baseline model without using DSR across three MARL environments, including LBF, RWARE, and SMAC.
Unlike traditional methods that require manually finding the suitable sight range, DSR automatically identifies the optimal sight range for each environment.
In addition, DSR accelerates the training process by gradually shifting the sight range from a smaller to a large one.
It further provides interpretability by offering insights into the selected sight ranges.
Finally, the method can be seamlessly integrated with multiple MARL algorithms without requiring algorithm-specific modifications.

Future work could explore how DSR generalizes to environments with different domains.  
For more complex environments, such as those with continuous observation spaces, the design of the meta-controller may require further adjustments. 
One possible approach is to discretize continuous sight ranges into representative discrete sets for selection.

The insights provided by DSR can also help sensor design in practical applications by balancing sight range, performance, and cost.  
In addition, in our work, all agents share the same sight range, but future research could investigate using different sight ranges for individual agents, which is particularly relevant for heterogeneous agent settings where agents have different roles or capabilities.  
In scenarios with a large number of options, the meta-controller may require further modifications to handle this issue effectively.  
Furthermore, deeper exploration is needed to understand how the structure of an environment influences the optimal sight range, providing new insights into the relationship between observation and environment dynamics.

\section*{Acknowledgement}

This research is partially supported by the National Science and Technology Council (NSTC) of the Republic of China (Taiwan) under Grant Number NSTC 113-2221-E-A49-127, NSTC 113-2221-E-A49-128, NSTC 113-2634-F-A49-004, and NSTC 113-2221-E-001-009-MY3.
The authors would like to thank Kuo-Hao Ho and Chiu-Chou Lin for some discussions in the early stages. 

\bibliographystyle{ACM-Reference-Format} \balance
% \bibliography{main}
\input{main.bbl}

\clearpage

\appendix
\renewcommand{\thesection}{\Alph{section}}
\renewcommand{\thesubsection}{\Alph{section}.\arabic{subsection}}

 \section{Environment Details}

\subsection{Level-Based Foraging (LBF)}

Level-Based Foraging (LBF) \cite{albrecht_GametheoreticModel_2013, papoudakis_BenchmarkingMultiAgent_2021} 
is a grid-based multi-agent environment designed to test multi-agent algorithms' ability to learn cooperation under different conditions. 
The environment includes various configurations, such as different numbers of agents, grid sizes, and food items. 
Some tasks require all agents to cooperate simultaneously to load an item. 
LBF has become a popular benchmark for evaluating multi-agent reinforcement learning (MARL) algorithms.

\textbf{Environment ID} 
In the Level-Based Foraging (LBF) environment, 
the naming convention for the environment ID typically reflects different properties of the environment, 
such as map size, the number of agents, the number of food items, and the observation range. 
The map size is usually denoted as ``NxN,'' 
indicating an N by N grid. 
For example, 8x8 refers to a grid of size 8x8. 
The number of agents (p) is indicated after the letter ``p,'' 
such as 2p, meaning 2 agents. 
The number of food items (f) is similarly denoted, 
with 2f, meaning there are 2 food items. 
The sight range (s) is also specified, 
such as 6s representing a sight range of 6. 
If the environment requires all agents to cooperate to complete the task (loading a food item), 
``coop'' indicates the cooperative mode. 
For instance, the environment ID 6s-10x10-4p-4f-coop represents a 10x10 map, 
with a sight range of 6, 
4 agents, 
4 food items, 
and a cooperative mode.

\textbf{Action Space} 
The action space includes the following actions: 
No Operation, 
move up, 
move down, 
move left, 
move right, 
and Load.

\textbf{Reward} 
The reward is proportional to the fraction of food collected relative to the total food score on the map.

\textbf{Sight Range Description} 
The sight range in LBF is represented visually, 
with a sight range of 1, indicating visibility within a 3x3 area around the agent. 
When the sight range is 2, 
the view expands to a 5x5 area.

\textbf{Observation} 
The length of the observation is based on the number of objects (agents and food) on the map, 
multiplied by the size of each object. 
If an object is outside the agent’s sight range, 
its value is set to a default. 
If it is within the sight range, 
its value is displayed.

\textbf{Termination} 
The episode terminates when all food has been collected 
or when the step limit of 50 steps is reached.

\subsection{Multi-Robot Warehouse (RWARE)}

The Multi-Robot Warehouse (RWARE) environment is a multi-agent, 
partially observable cooperative task where robots work together to move goods in a grid-world warehouse. 
The agents must locate and deliver requested shelves to designated goal locations 
and then return the shelves to empty positions before continuing with the next delivery. 
Once a shelf is successfully moved to the goal position, 
new requested shelves are randomly generated. 
The more deliveries the agents complete within the step limit, 
the higher the team’s total reward.

\textbf{Environment ID} 
The environment ID in RWARE includes indicators for map size, 
such as ``tiny'' and ``small.'' 
In the ``tiny'' setting, 
there are 2 columns of shelves, 
while in ``small,'' 
there are 5 columns, 
indicating a larger map. 
The number of requested shelves equals the number of agents (denoted by ``ag''). 
We further add ``s'' to show different default sight range settings.

\textbf{Actions} 
Each agent can choose from the following actions: 
do nothing, 
move forward, 
rotate left, 
rotate right, 
and pick up or drop off a shelf.

\textbf{Reward} 
Completing a series of actions, 
including locating a shelf, 
moving it to the goal, 
and returning it, 
is necessary to obtain a reward. 
Each completed delivery yields +1 reward.

\textbf{Sight Range Description} 
The sight range in RWARE is similar to LBF, 
with a sight range of 1 indicating visibility within a 3x3 area around the agent. 
A sight range of 2 expands the view to a 5x5 area.

\textbf{Observation} 
The observation includes information about each grid within the agent's visible range, 
such as whether it contains an agent, a shelf, or status information. 
To avoid inconsistencies in observation size, 
if a smaller sight range is selected, 
the grids outside this range are masked to 0. 
For example, 
if the original sight range is 5s and the meta-controller selects 3s, 
the corresponding grids outside 3s will be masked in the observation. 
Note that LBF and SMAC do not have this issue, 
as their observation size is determined by the total number of objects, 
not by the sight range.

\textbf{Termination} 
The episode terminates when the step limit of 500 steps is reached.

\subsection{StarCraft Multi-Agent Challenge (SMAC)}

The StarCraft Multi-Agent Challenge (SMAC) \cite{samvelyan_StarCraftMultiAgent_2019} is a widely used benchmark for multi-agent reinforcement learning (MARL), based on the real-time strategy game StarCraft II.  
SMAC focuses specifically on micromanagement scenarios, where each unit is controlled by an independent agent that acts based on local observations.  
The environment introduces significant challenges related to partial observability, decentralized execution, and the coordination of multiple agents.  
In SMAC, allied units, controlled by learned agents, face off against enemy units directed by the game's built-in AI.  
The scenarios vary in terms of the number, type, and position of units, as well as the terrain.  
The primary objective is to maximize the win rate by effectively managing the actions and coordination of individual units.  

The actions available to the agents include move, attack, stop, and other commands, depending on the unit types.  
This environment provides a default dense reward function, which is related to the damage dealt and the death of units.  
Different scenarios in SMAC have varying time limits, determining when an episode ends.  
The sight range in SMAC is defined as the radius of a circle centered around each agent.  
The default sight range is 9, while the shooting range is 6.  

The observation includes information about allied and enemy units within the agent's sight range, such as distance, relative coordinates, health, shield, and unit type.  
SMAC provides global states for centralized learning, which contains information about all units on the map, such as their absolute positions.  
Considering more general cases where the environment does not provide a state, it is common to concatenate the agents' observations to construct a state.  
We focus on the complexity of observations in this study, thus using a version of SMAC where observations are combined into a state.  
Notably, environments like LBF and RWARE, by default, construct the state by combining agent observations.

One part of the original observation in SMAC is the relative distance between other units and the agent itself. This value is normalized by the sight range. In the implementation, to accommodate different possible sight ranges, the normalization is consistently done by dividing by 9.

In SMAC, the environment IDs follow a structured naming convention that reflects the units involved in each scenario.  
The format is typically: \textit{<Ally units>\_vs\_<Enemy units>}.  
The first part of the ID describes the composition of the allied units, using abbreviations such as \textit{s} for Stalkers, \textit{z} for Zealots, \textit{m} for Marines, and \textit{c} for Colossus.  
Numbers preceding these abbreviations indicate the number of units of that type.  
The \textit{vs} separator indicates a battle between the allied and enemy units, with the second part of the ID describing the enemy units in the same format.  
If the ID does not contain the \textit{vs} separator, some cases indicate identical lineups on both sides, while other situations may also occur. Table \ref{tab:smac-name} provides some examples.

\begin{table*}
    \caption{The SMAC challenges.}
    \centering
    \begin{tabular}{ccc}
    \toprule
        Map Name    & All Units & Enemy Units \\
    \midrule
        5m\_vs\_6m    & 5 Marines  & 6 Marines \\
        8m\_vs\_9m    & 8 Marines & 9 Marines \\
        10m\_vs\_11m  & 3 Stalkers + 5 Zealots & 3 Stalkers + 5 Zealots \\
        3s\_vs\_5z    & 3 Stalkers & 5 Zealots\\
        3s5z        & 3 Stalkers + 5 Zealots &3 Stalkers + 5 Zealots \\
        MMM2        & 1 Medivac + 2 Marauders + 7 Marines & 1 Medivac + 3 Marauders + 8 Marines\\
    \bottomrule
    \end{tabular}
    \label{tab:smac-name}
\end{table*}

\subsection{SMAC Dynamic Team Composition}

SMAC Dynamic Team Composition (SMAC-DT) maps are dynamic variations of the standard StarCraft II micromanagement tasks used in \cite{shao_ComplementaryAttention_2023, iqbal_RandomizedEntitywise_2021}, where both the number and types of units (agents) can vary between episodes. 
These maps feature symmetrical battles between two teams of units, with the configuration of each team changing randomly, creating a wide variety of scenarios for training multi-agent systems. 
Following \cite{shao_ComplementaryAttention_2023, iqbal_RandomizedEntitywise_2021}, we use the entity-based SMAC environment.
In our training, each task randomly selects a combination of 3 to 5 agents for each battle.

\section{Experiment Details}

LBF, RWARE, and SMAC environments each conducted 100 testing episodes at every test point during training. 
The test intervals were set as follows: 50,000 steps for LBF, 500,000 steps for RWARE, and 100,000 steps for SMAC.
For the SMAC-DT experiment, we utilized CAMA's code base and followed its settings, running 160 testing episodes with a test interval of 50,000 steps. The figure for SMAC-DT shows the recent test win rate reported in the log, with a log interval of 10,000 steps.
The experiments were run on a machine with an E5-2683 v3 @ 2.00GHz CPU and 4 GTX 1080 Ti GPUs.

\subsection{Hyperparameters}

\textbf{General settings.}
For all environments, including LBF, RWARE, SMAC, and SMAC-DT, the discount factor (\(\gamma\)) is set to 0.99, and gradient norm clipping is applied with a value of 10. The batch size is 32 for Q-learning-based algorithms (QMIX, IQL, VDN, and IM-Qatten) and 10 for IPPO and MAPPO. 

\textbf{Hyperparameters of UCB.}
For all DSR's UCB instances, we use the default settings of $c=2$ and $w=5000$. 
Regarding the episode return, LBF’s return naturally falls within the range of 0 to 1, while for RWARE, most of the returns are below 1 during the early stages of training. 
In contrast, SMAC often has returns greater than 1 (figure \ref{fig:smac-allResults-DsrVsOrigianl-Return}), so we normalize the returns of SMAC and SMAC-DT by dividing by 20 during UCB updates to approximately bring the values within the range of 0 to 1. No specific normalization is applied for LBF or RWARE.

\textbf{Hyperparameters of each algorithm.}
The detailed hyperparameters for each algorithm and domain are listed below.
\begin{itemize}
    \item Table \ref{tab:hyper-qmix-basic}: the hyperparameters of QMIX across different domains
    \item Table \ref{tab:hyper-IMQatten-basic}: the hyperparameters of IM-Qatten for SMAC-DT
    \item Table \ref{tab:hyper-IQLAndVDN}: the hyperparameters of both IQL and VDN in LBF and RWARE.
    \item Table \ref{tab:hyper-IPPO}: the hyperparameters of IPPO in LBF and RWARE.
    \item Table \ref{tab:hyper-MAPPO}: the hyperparameters of MAPPO in LBF and RWARE.
\end{itemize}
\vspace{-5pt}

\subsection{Details of SMAC-DT Experiments}

Our goal is to compare different observation selection mechanisms.
To ensure a fair comparison of observation selection methods without any additional communication mechanisms or information, the global coach is disabled for the CAMA method. 
We use the source code of CAMA and incorporate our method, following its entity-based environment setup for training in the dynamic team composition tasks of SMAC. 
The baseline is IM-Qatten (Qatten \cite{yang_QattenGeneral_2020} with an inverse model \cite{pathak_CuriositydrivenExploration_2017}), and we compare the performance of using DSR with CAMA's attention-based ranking selection mechanism.

\begin{table}
    \centering
    \caption{Hyperparameters for QMIX across different domains.}
    \begin{tabular}{lccc}
        \toprule
                                  &  LBF       & RWARE      & SMAC \\
        \midrule
        Hidden dimension          & 128        & 128        & 128  \\
        Learning rate             & 0.0003     & 0.0005      & 0.0005 \\
        Reward standardisation    & True       & True       & True \\
        Network type              & GRU        & GRU        & GRU \\
        Parameter sharing         & True       & True       & True \\
        Epsilon start             & 1          & 1          & 1 \\
        Epsilon finish            & 0.05       & 0.05       & 0.05 \\ 
        Evaluation epsilon        & 0.05       & 0.05       & 0.05 \\
        Epsilon anneal            & 200000     & 1000000    & 50000 \\
        Target update             & 200 (hard) & 200 (hard) & 200 (hard) \\
        Replay buffer size        & 5000       & 500        & 5000\\
        Parallel runner number    &  1         & 1          & 1 \\
        Mixing embed dimension    & 32         & 32         & 32\\
        Hypernet layers           & 2          & 2          & 2\\
        Hypernet embed dimension  & 64         & 64         & 64\\
        Optimizer                 & Adam       & Adam       & Adam \\
        \bottomrule
    \end{tabular}
    \label{tab:hyper-qmix-basic}
\end{table}

\begin{table}
    \centering
    \caption{Hyperparameters for IM-Qatten in SMAC-DT.}
    \begin{tabular}{lc}
        \toprule
                                  &  SMAC-DT     \\
        \midrule
        Epsilon start             & 1        \\
        Epsilon finish            & 0.05     \\ 
        Epsilon anneal            & 200000   \\
        Evaluation epsilon        & 0     \\        
        Replay buffer size           & 5000       \\
        Parallel runner number       &  8         \\
        Number of multi-head attention heads &  4  \\
        Dimension of Attention embedding in local agent & 128 \\
        Mixing embedding dimension & 32 \\
        RNN hidden dimension      & 64      \\
        Optimizer                 & RMSprop \\
        Learning rate             & 0.0005   \\
        Target update             & 200 (hard) \\
        Weight for $\mathcal{L}_{IM}$       & 0.005 \\       
        \bottomrule
    \end{tabular}
    \label{tab:hyper-IMQatten-basic}
\end{table}

\begin{table}
    \centering
    \caption{Hyperparameters for IQL and VDN in LBF and RWARE.}
    \begin{tabular}{lcc}
        \toprule
                                  &  LBF       & RWARE      \\
        \midrule
        Hidden dimension          & 128        & 128        \\
        Learning rate             & 0.0003     & 0.0005     \\
        Reward standardisation    & True       & True       \\
        Network type              & GRU        & GRU        \\
        Parameter sharing         & True       & True        \\
        Epsilon start             & 1          & 1         \\
        Epsilon finish            & 0.05       & 0.05       \\ 
        Evaluation epsilon        & 0.05       & 0.05       \\
        Epsilon anneal            & 200000     & 1000000   \\
        Target update             & 200 (hard) & 200 (hard)  \\
        Replay buffer size        & 5000       & 500      \\
        Parallel runner number    &  1         & 1          \\
        Optimizer                 & Adam       & Adam       \\
        \bottomrule
    \end{tabular}
    \label{tab:hyper-IQLAndVDN}
\end{table}

\begin{table}
    \centering
    \caption{Hyperparameters for IPPO in LBF and RWARE.}
    \begin{tabular}{lcc}
        \toprule
                                  &  LBF        & RWARE      \\
        \midrule
        Hidden dimension          & 64          & 64        \\
        Learning rate             & 0.0003      & 0.0005     \\
        Reward standardisation    & False       & False       \\
        Network type              & GRU         & GRU        \\
        Parameter sharing         & True        & True        \\
        Entropy coefficient       & 0.001       & 0.001\\
        Q n-step                  & 5           & 10 \\
        Epochs                    & 4           & 4 \\
        PPO clip                  & 0.2         & 0.2 \\
        Target update             & 200 (hard)  & 0.01 (soft)   \\
        Parallel runner number    &  10         & 10          \\
        Optimizer                 & Adam        & Adam       \\
        \bottomrule
    \end{tabular}
    \label{tab:hyper-IPPO}
\end{table}

\begin{table}
    \centering
    \caption{Hyperparameters for MAPPO in LBF and RWARE.}
    \begin{tabular}{lcc}
        \toprule
                                  &  LBF        & RWARE      \\
        \midrule
        Hidden dimension          & 64          & 64        \\
        Learning rate             & 0.0003      & 0.0005     \\
        Reward standardisation    & False       & False         \\
        Network type              & GRU         & GRU        \\
        Parameter sharing         & True        & True        \\
        Entropy coefficient       & 0.001       & 0.001 \\
        Q n-step                  & 5           & 10 \\
        Epochs                    & 4           & 4 \\
        PPO clip                  & 0.2         & 0.2 \\
        Target update             & 0.01 (soft) & 0.01 (soft)  \\
        Parallel runner number    &  10         & 10          \\
        Optimizer                 & Adam        & Adam       \\
        \bottomrule
    \end{tabular}
    \label{tab:hyper-MAPPO}
\end{table}

\section{Full Experiment Results}
 
\subsection{QMIX across Different Domains}

\begin{figure*}
    \centering
    \includegraphics[width=1\linewidth]{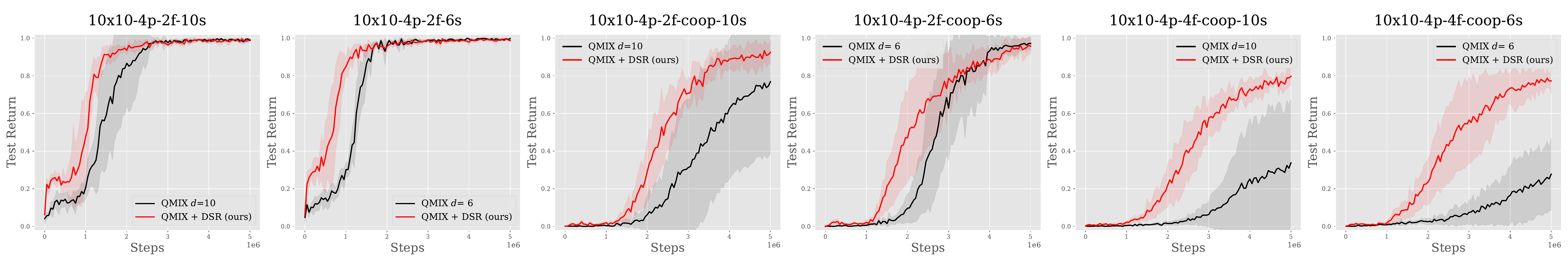}
    \caption{Full results of QMIX in LBF: w/o DSR (original sight range) vs. w/ DSR}
    \label{fig:lbf-allQMIX-originalVsDSR}
\end{figure*}

\begin{figure*}
    \centering
    \includegraphics[width=1\linewidth]{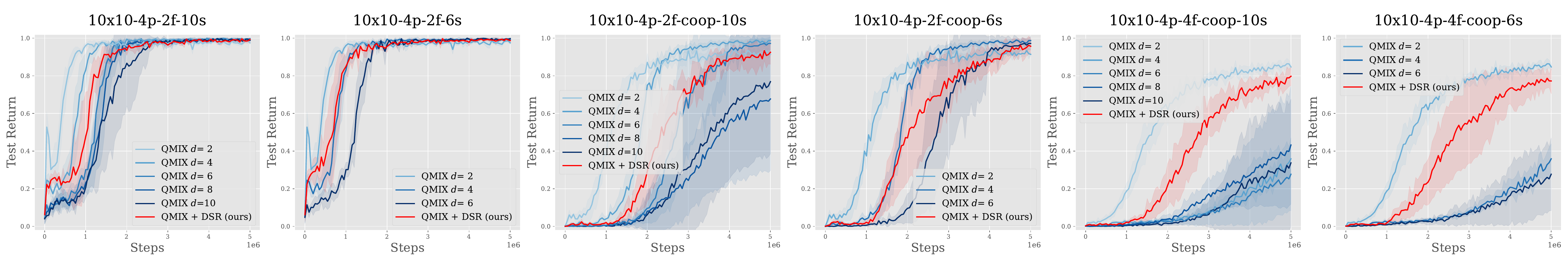}
    \caption{Full results of QMIX in LBF: Training with fixed sight ranges vs. DSR. Note that the largest fixed sight range corresponds to the original sight range for that task.}
    \label{fig:lbf-allQMIX-allPuresVsDSR}
\end{figure*}

\begin{figure*}
    \centering
    \includegraphics[width=1\linewidth]{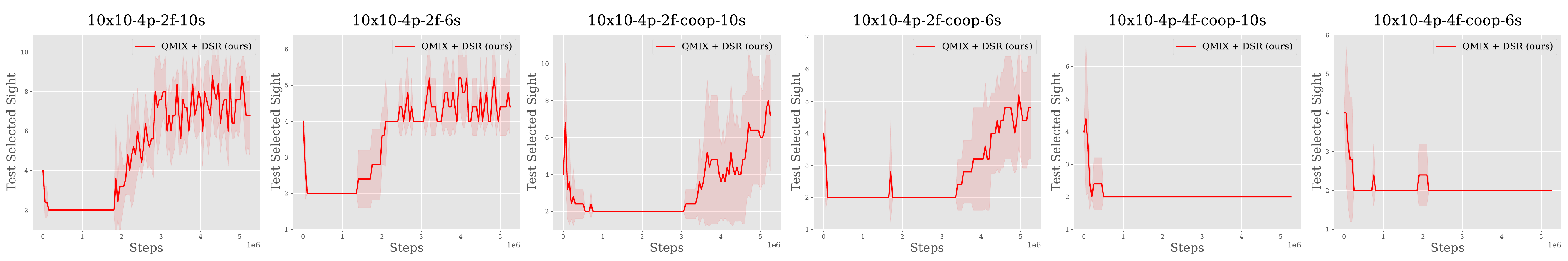}
    \caption{Full results of QMIX in LBF}
    \label{fig:lbf-allQMIX-selectedSights}
\end{figure*}

\begin{figure*}
    \centering
    \includegraphics[width=1\linewidth]{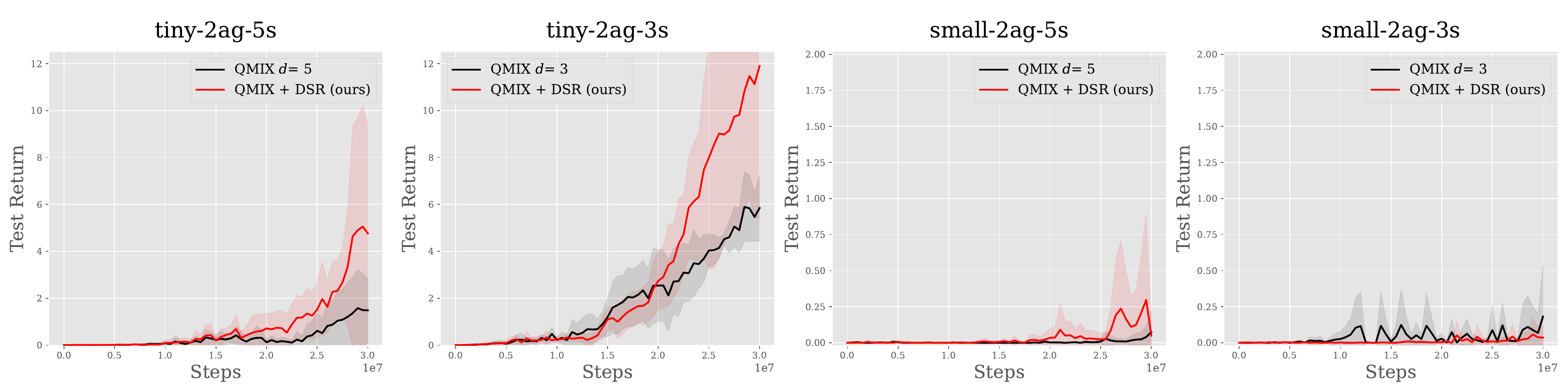}
    \caption{Full results of QMIX in RWARE: w/o DSR (original sight range) vs. w/ DSR}
    \label{fig:rware-allQMIX-originalVsDSR}
\end{figure*}

\begin{figure*}
    \centering
    \includegraphics[width=1\linewidth]{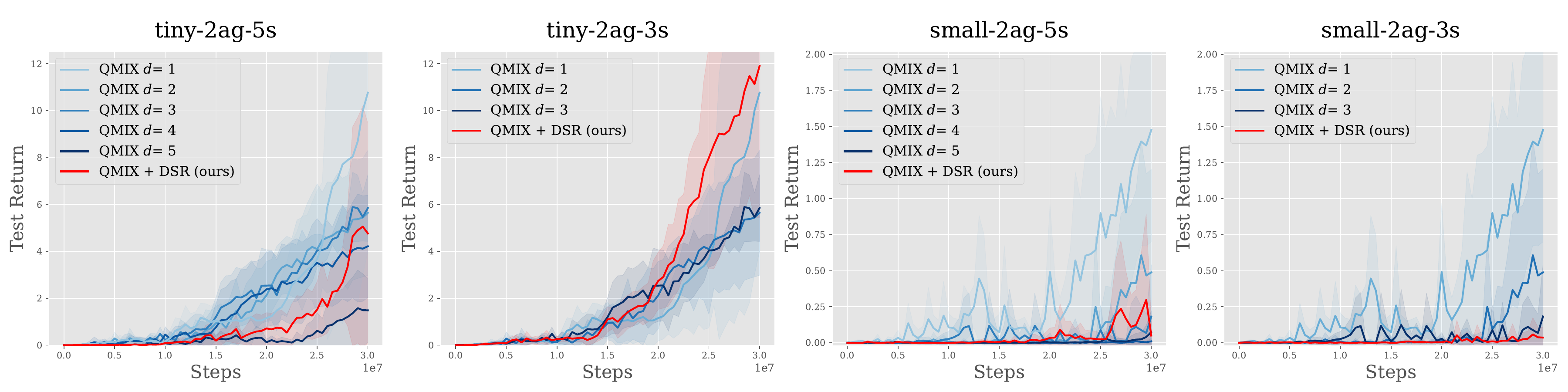}
    \caption{Full results of QMIX in RWARE: Training with fixed sight ranges vs. DSR. Note that the largest fixed sight range corresponds to the original sight range for that task.}
    \label{fig:rware-allQMIX-allPuresVsDSR}
\end{figure*}

\begin{figure*}
    \centering
    \includegraphics[width=1\linewidth]{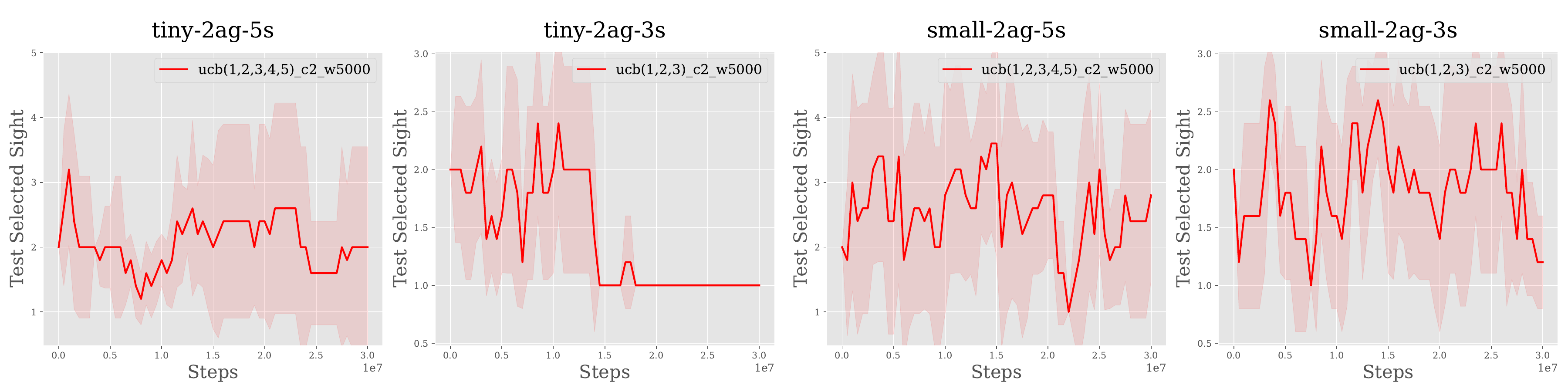}
    \caption{Full results of QMIX in RWARE}
    \label{fig:rware-allQMIX-selectedSights}
\end{figure*}

Figures \ref{fig:lbf-allQMIX-originalVsDSR}, \ref{fig:lbf-allQMIX-allPuresVsDSR}, and \ref{fig:lbf-allQMIX-selectedSights} display the complete results of QMIX on LBF tasks.
Figures \ref{fig:rware-allQMIX-originalVsDSR}, \ref{fig:rware-allQMIX-allPuresVsDSR}, and \ref{fig:rware-allQMIX-selectedSights} show the complete results of QMIX on RWARE tasks.

\begin{figure*}
    \centering
    \includegraphics[width=1\linewidth]{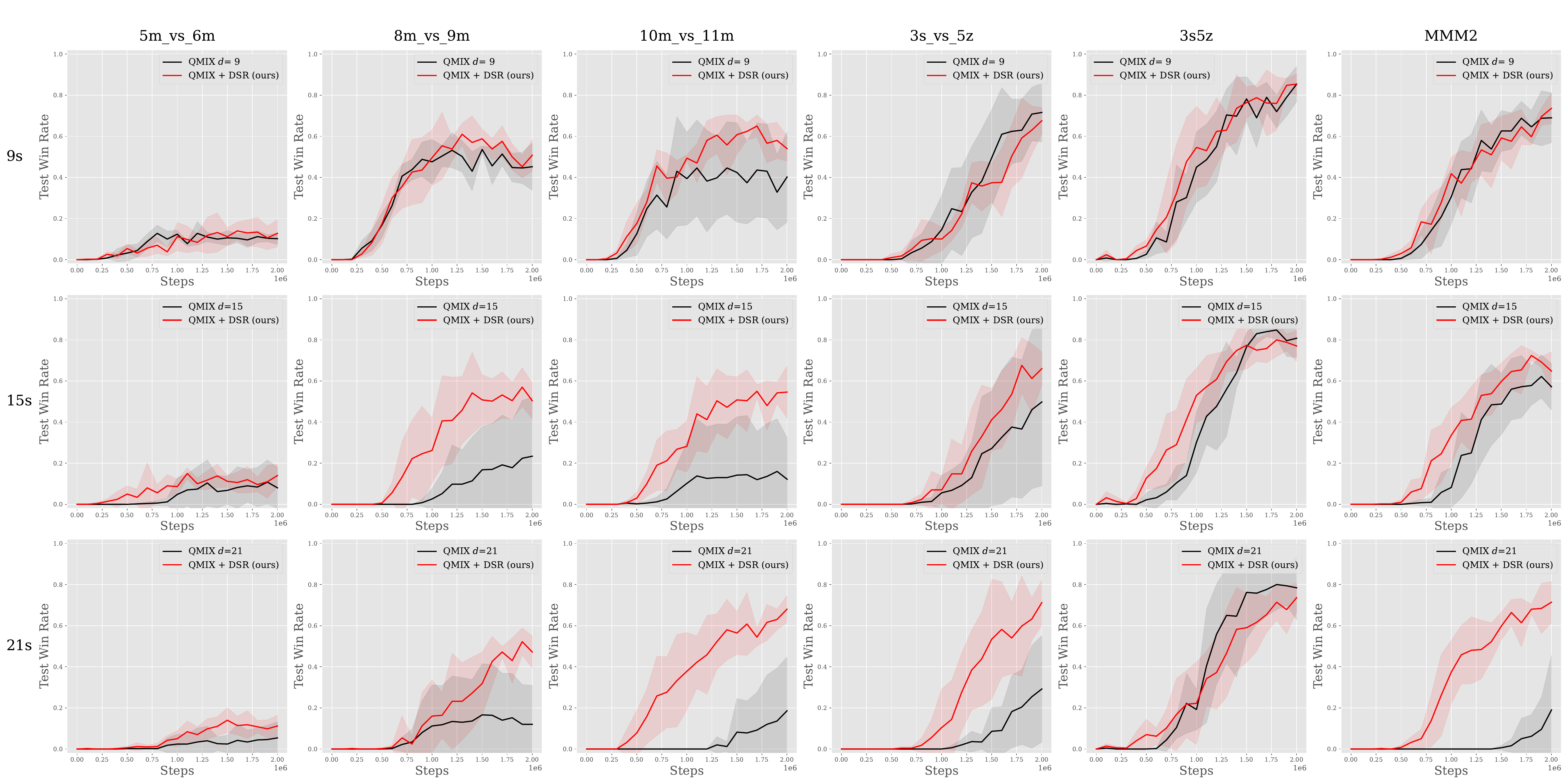}
    \caption{Test win rate of with DSR and without DSR in different maps and original sight ranges in SMAC.}
    \label{fig:smac-allResults-DsrVsOrigianl-win}
\end{figure*}

\begin{figure*}
    \centering
    \includegraphics[width=0.8\linewidth]{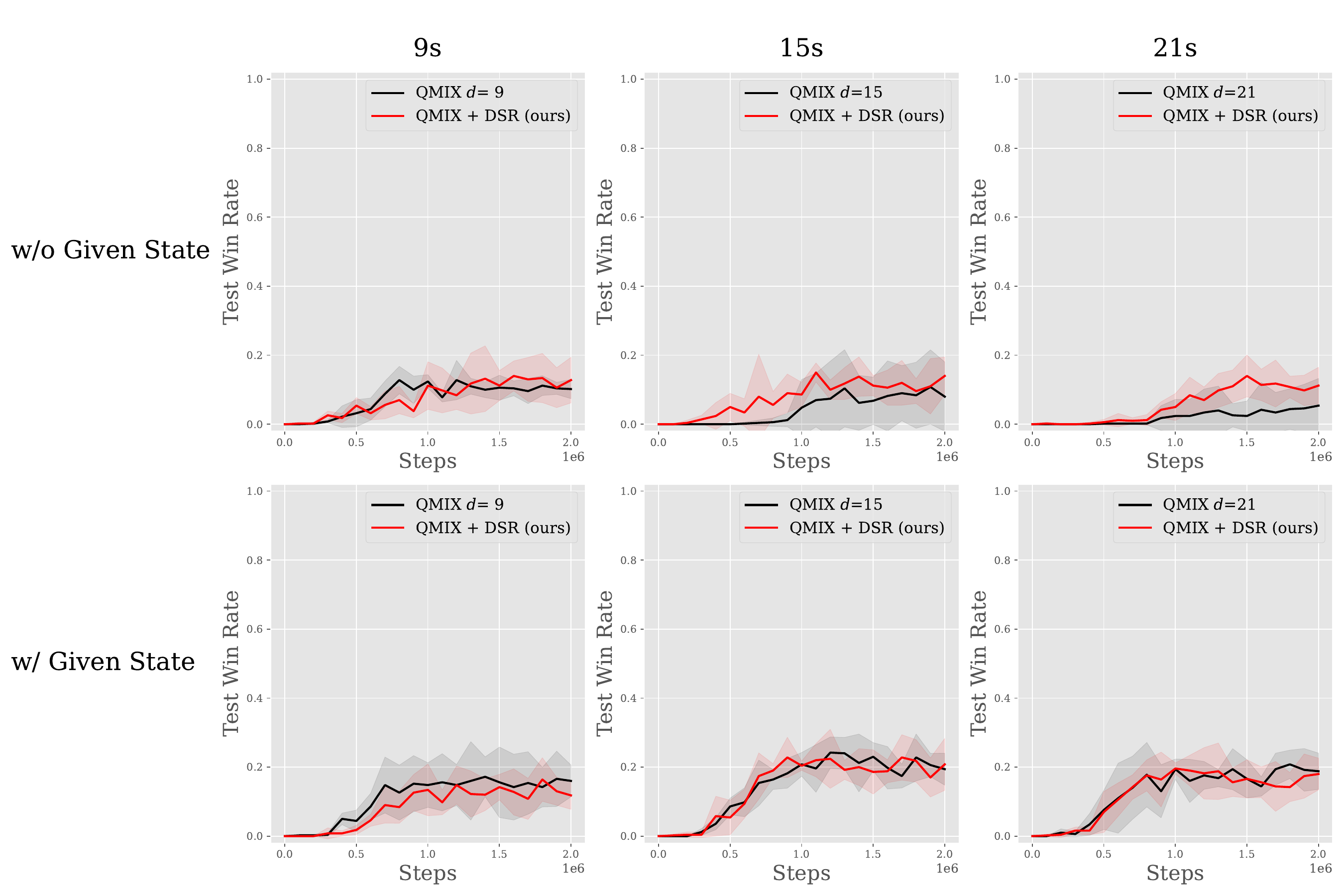}
    \caption{Full results of SMAC 5m\_vs\_6m.}
    \label{fig:smac-allResults-5mVs6m}
\end{figure*}

\begin{figure*}
    \centering
    \includegraphics[width=0.8\linewidth]{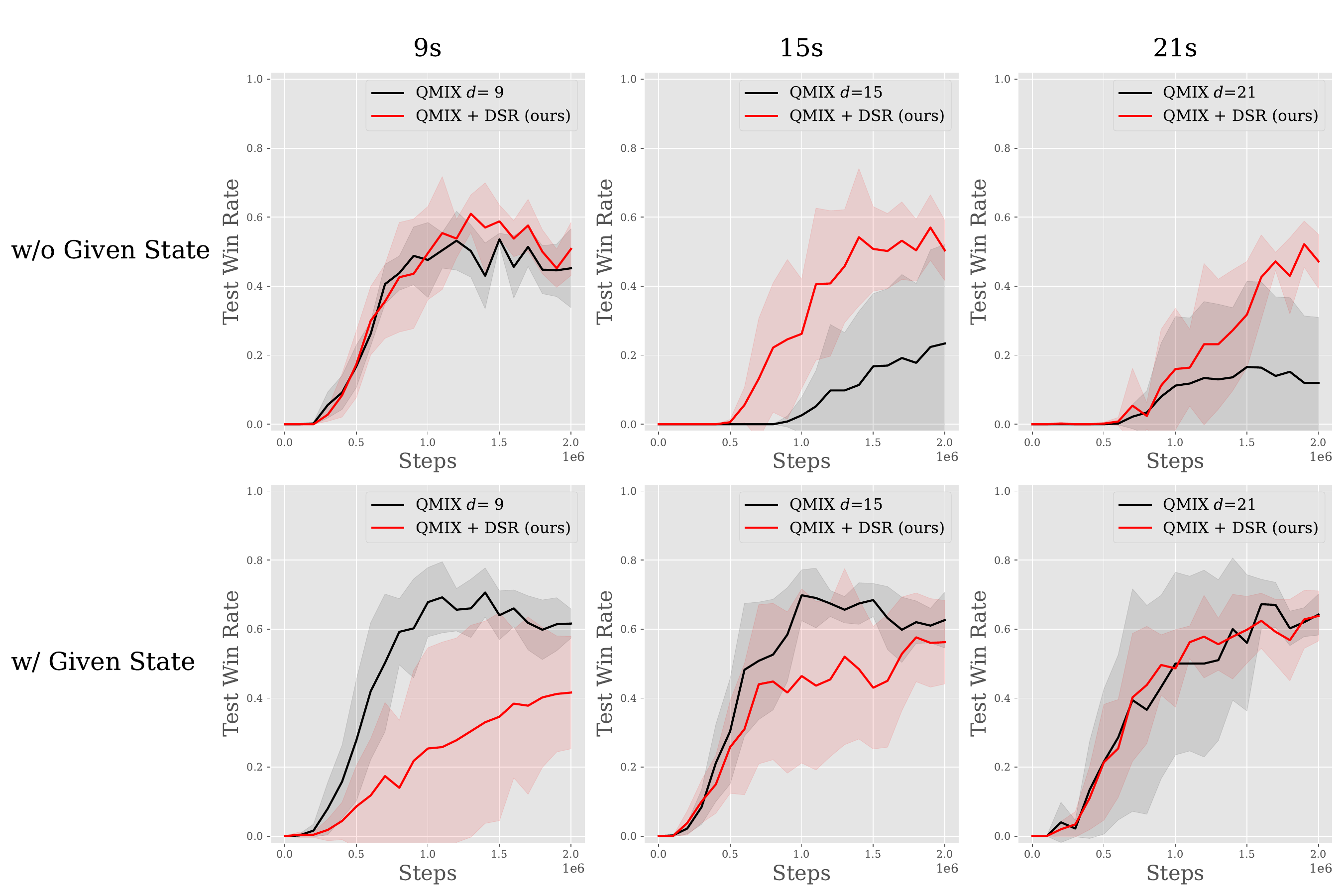}
    \caption{Full results of SMAC 8m\_vs\_9m.}
    \label{fig:smac-allResults-8mVs9m}
\end{figure*}

\begin{figure*}
    \centering
    \includegraphics[width=0.8\linewidth]{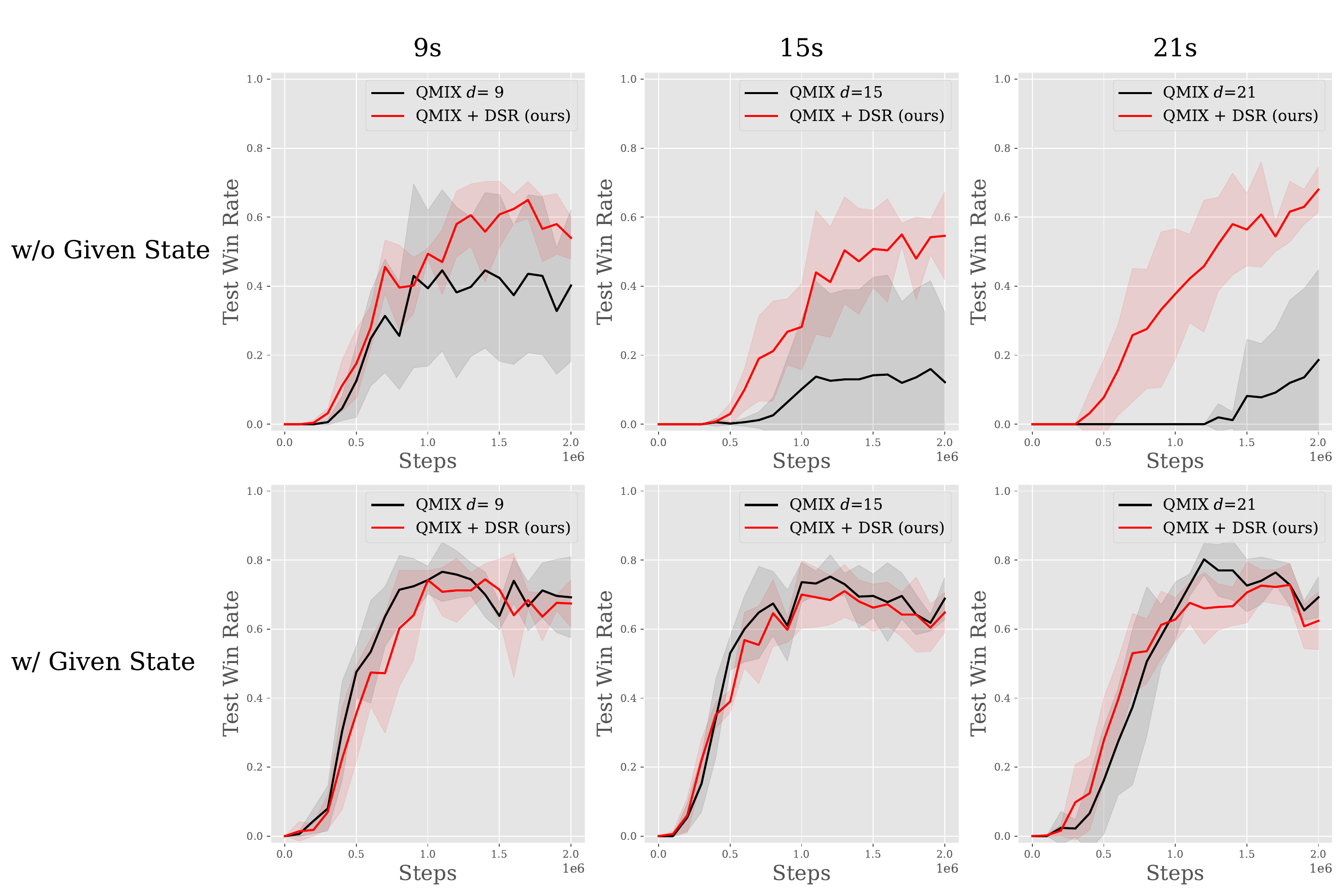}
    \caption{Full results of SMAC 10m\_vs\_11m.}
    \label{fig:smac-allResults-10mVs11m}
\end{figure*}

\begin{figure*}
    \centering
    \includegraphics[width=0.8\linewidth]{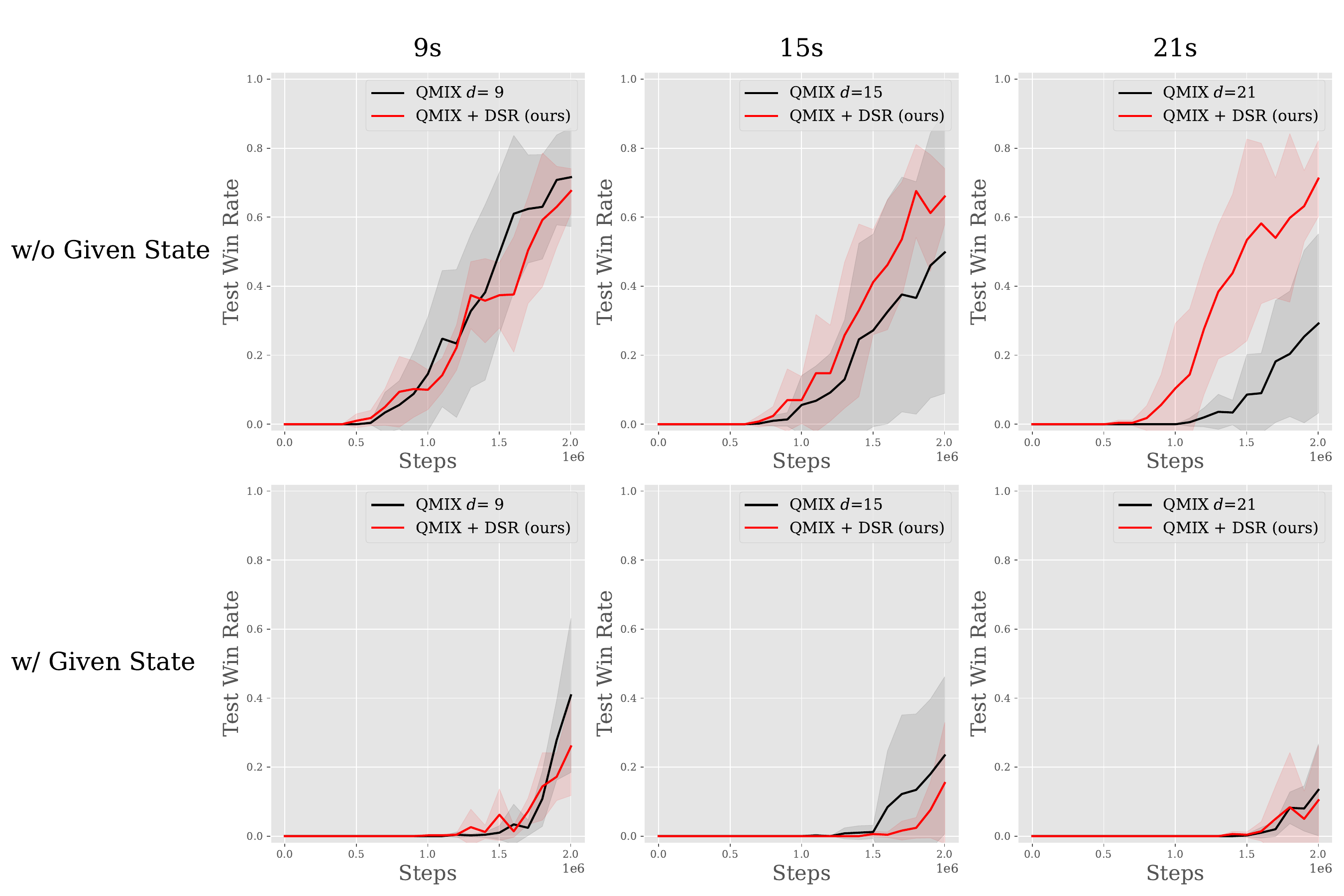}
    \caption{Full results of SMAC 3s\_vs\_5z.}
    \label{fig:smac-allResults-3s_vs_5z}
\end{figure*}

\begin{figure*}
    \centering
    \includegraphics[width=0.8\linewidth]{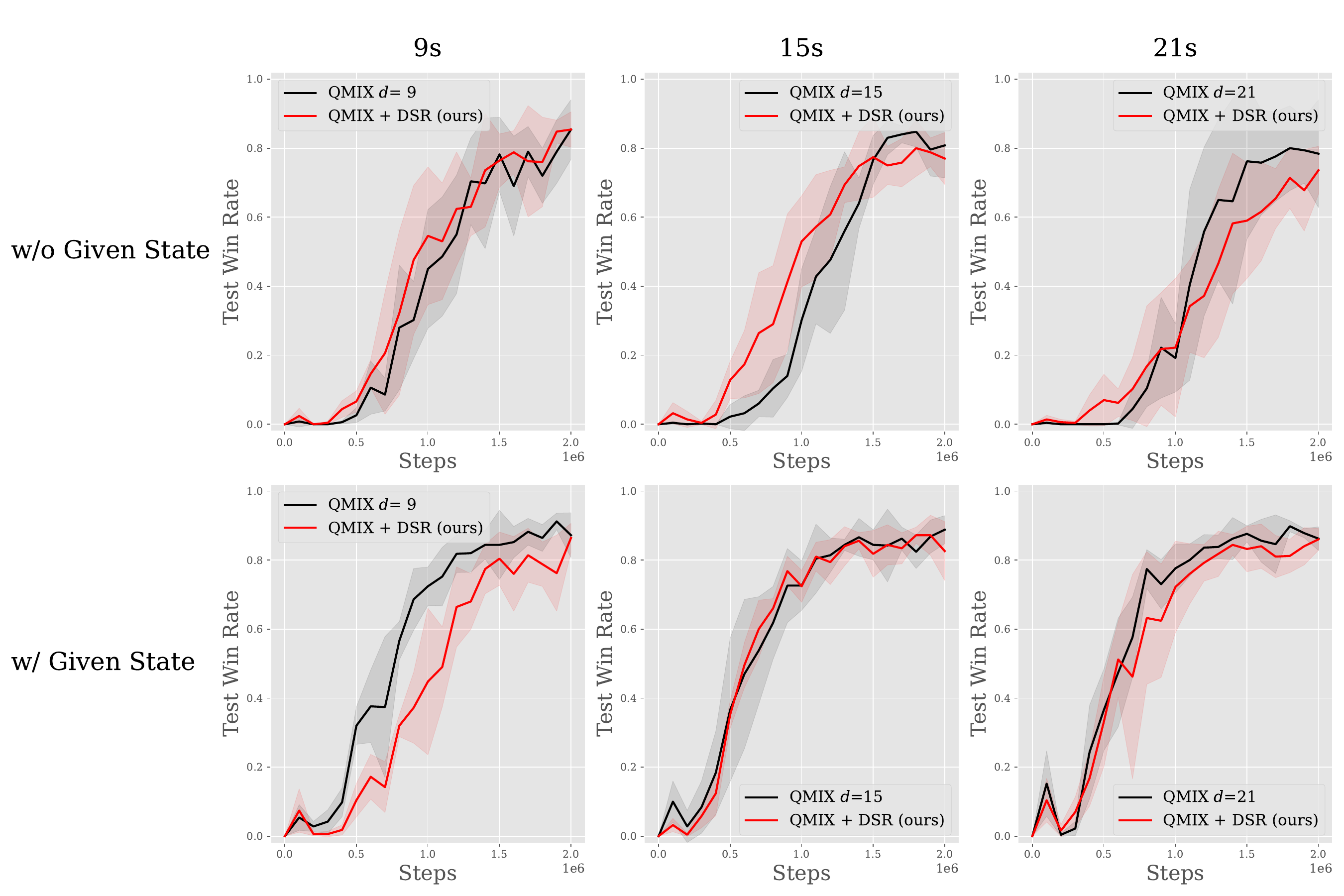}
    \caption{Full results of SMAC 3s5z.}
    \label{fig:smac-allResults-3s5z}
\end{figure*}

\begin{figure*}
    \centering
    \includegraphics[width=0.8\linewidth]{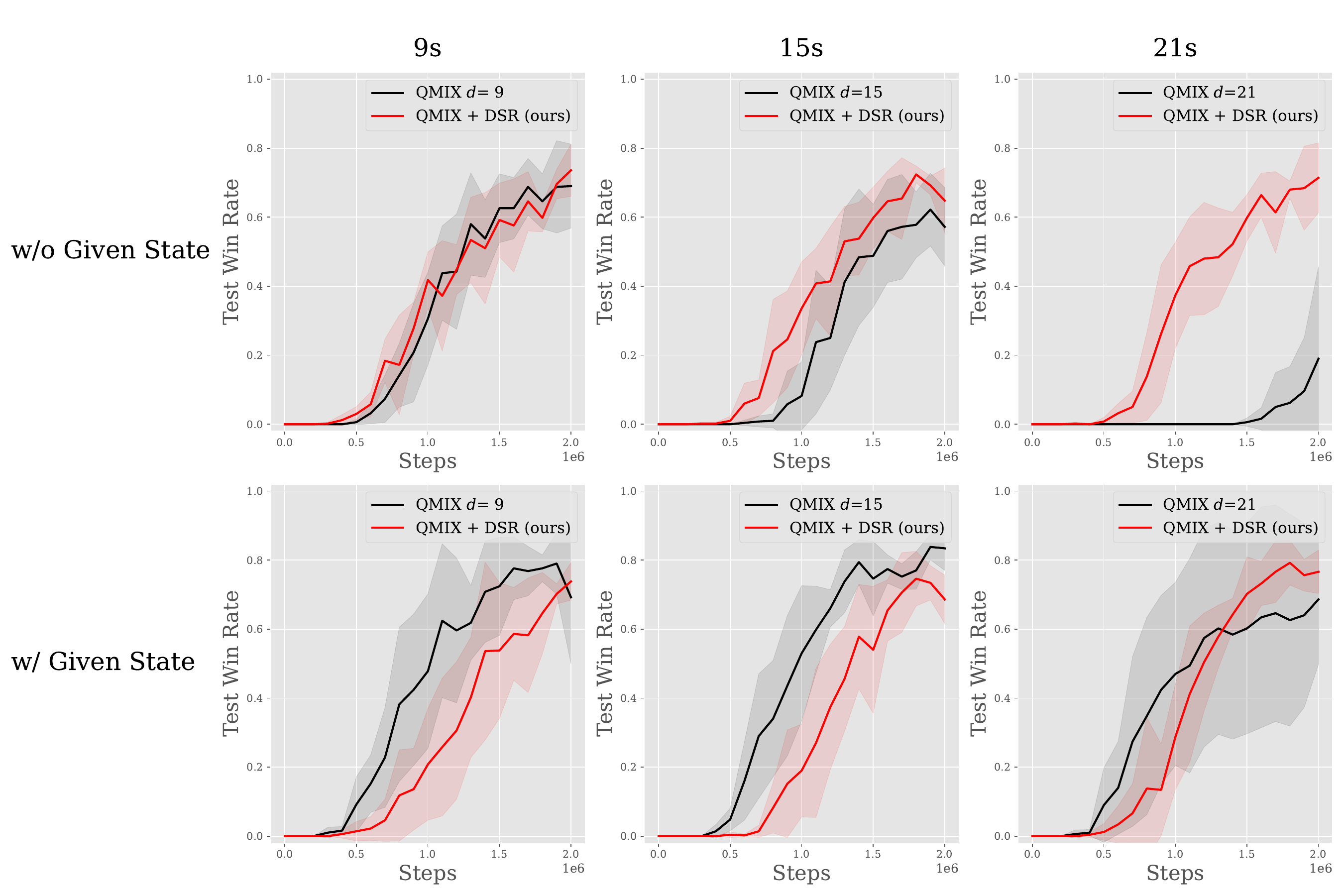}
    \caption{Full results of SMAC MMM2.}
    \label{fig:smac-allResults-MMM2}
\end{figure*}

\begin{figure*}
    \centering
    \includegraphics[width=1\linewidth]{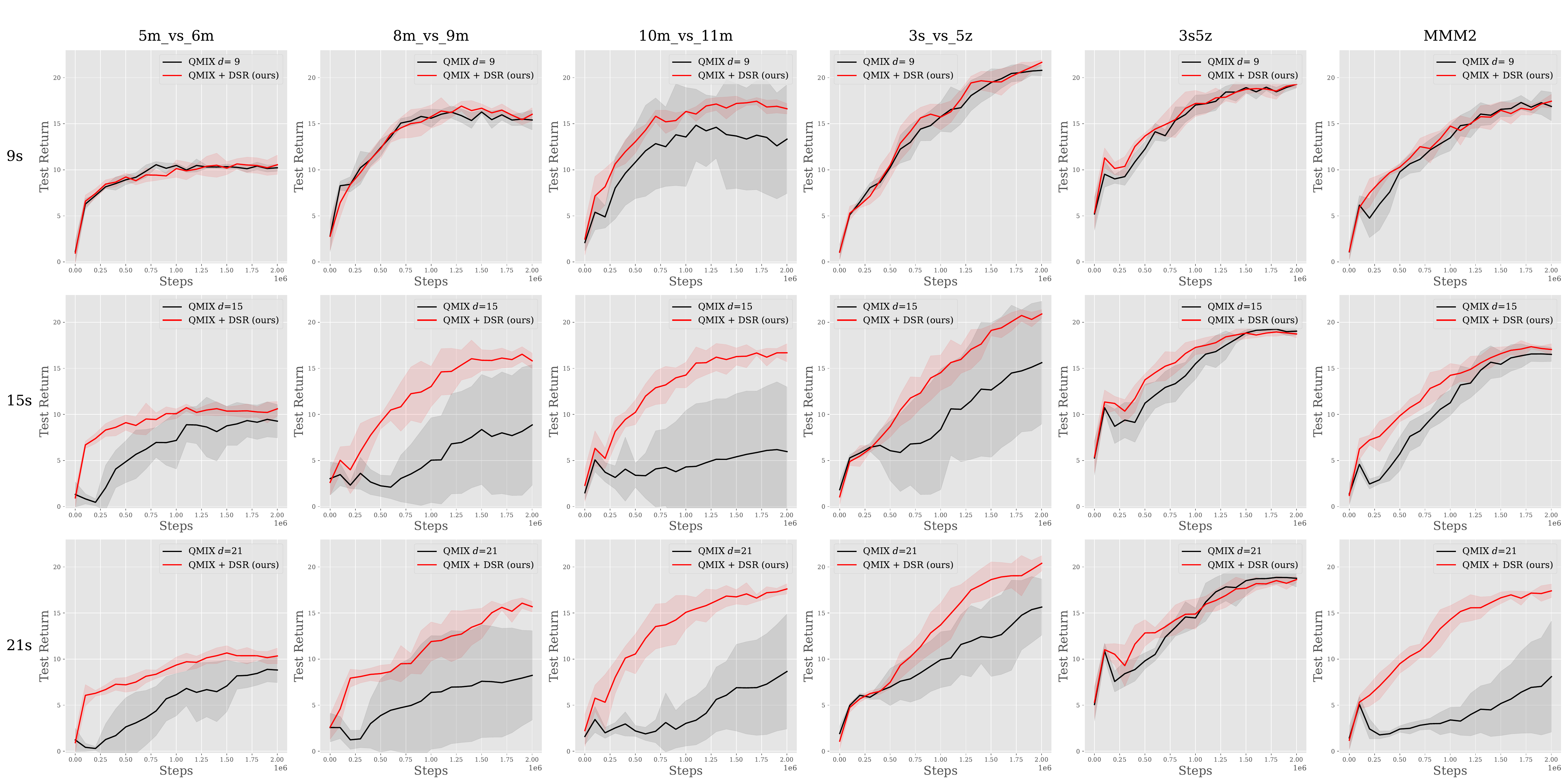}
    \caption{Test return of with DSR and without DSR in different maps and original sight ranges in SMAC.}
    \label{fig:smac-allResults-DsrVsOrigianl-Return}
\end{figure*}

\begin{figure*}
    \centering
    \includegraphics[width=1\linewidth]{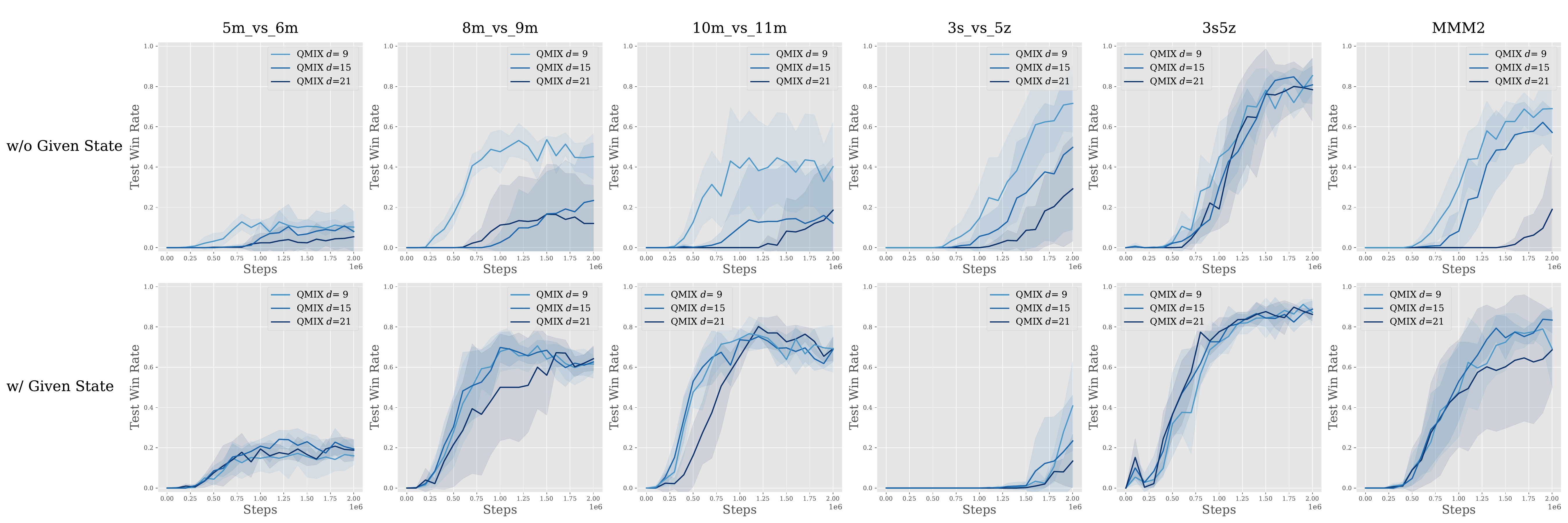}
    \caption{Test win rate across different maps in SMAC, various sight ranges ($d$), and whether using the global state or combining observations as state input.}
    \label{fig:smac-allResults-MapVsStateVsPures-win}
\end{figure*}

\begin{figure*}
    \centering
    \includegraphics[width=1\linewidth]{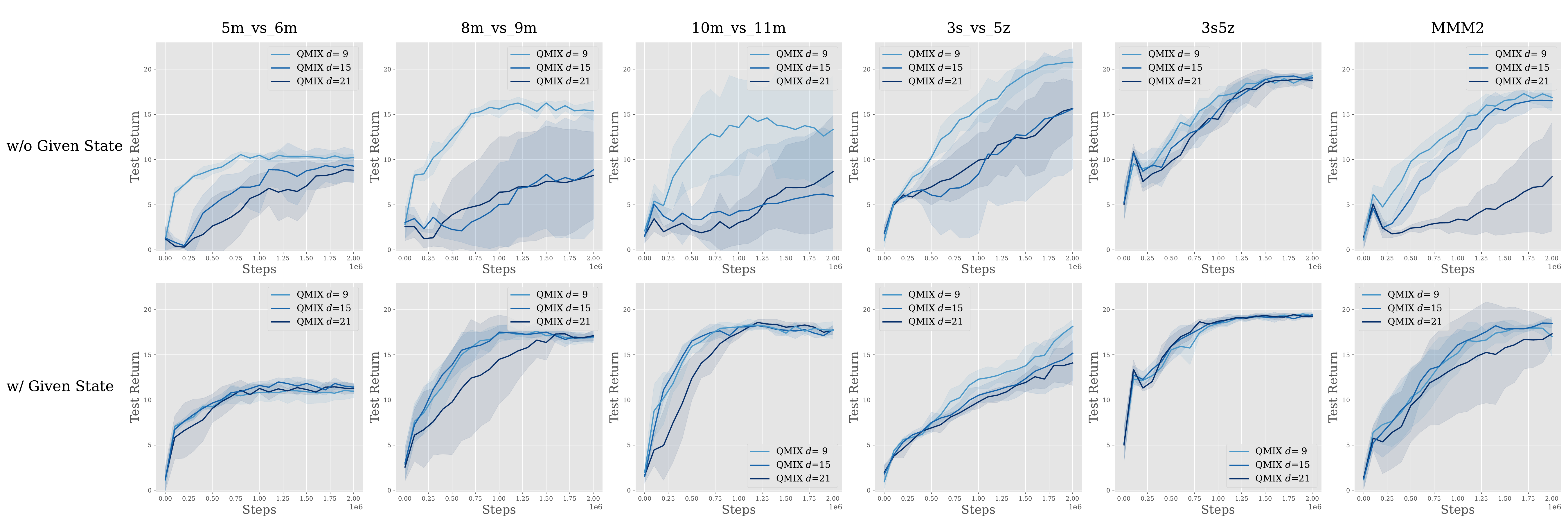}
    \caption{Test return across different maps in SMAC, various sight ranges ($d$), and whether using the global state or combining observations as state input.}
    \label{fig:smac-allResults-MapVsStateVsPures-Return}
\end{figure*}

\begin{figure*}
    \centering
    \includegraphics[width=1\linewidth]{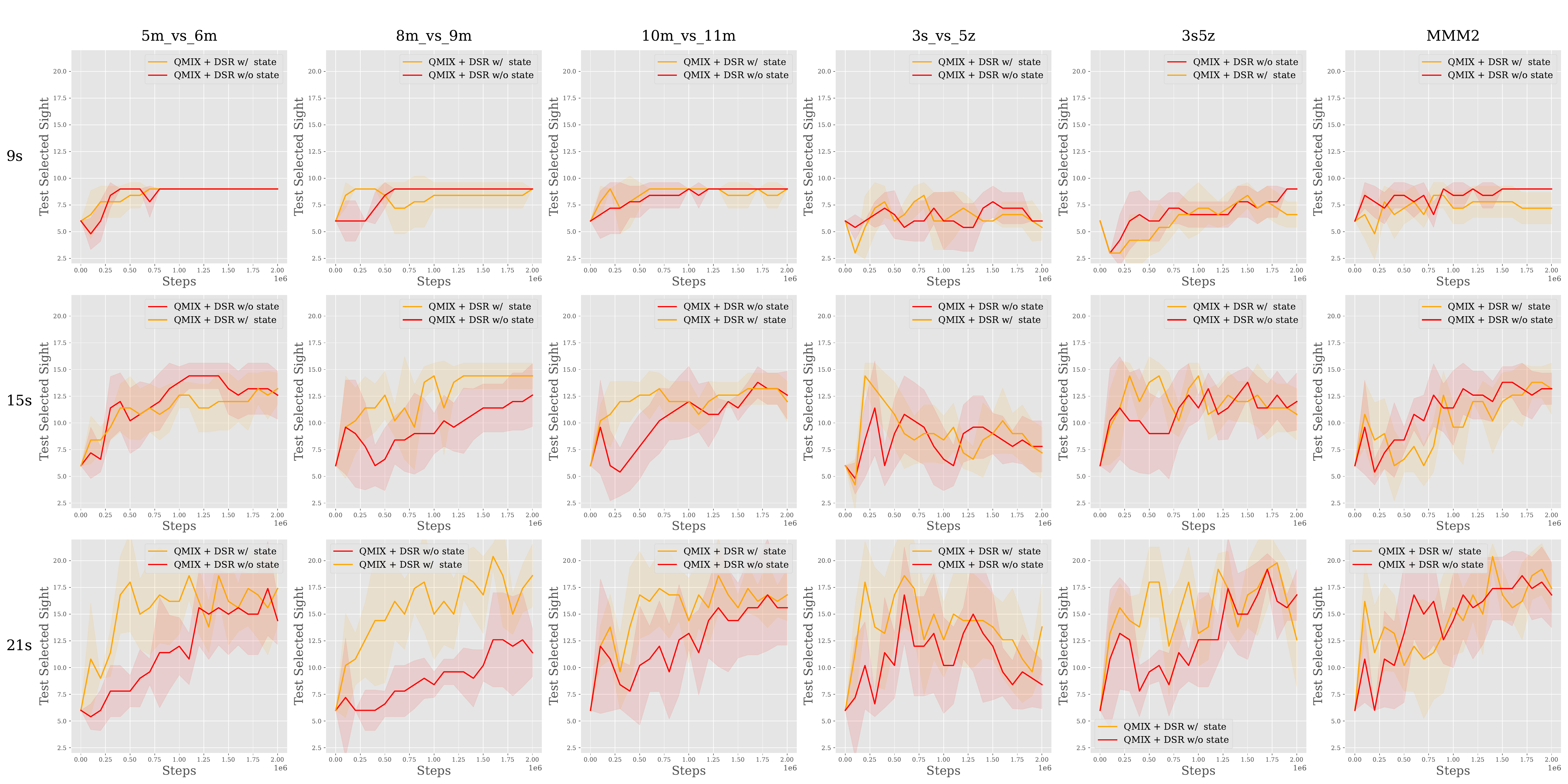}
    \caption{Test selected sights across different maps in SMAC, various sight ranges ($d$), and whether using the global state or combining observations as state input.}
    \label{fig:smac-allResults-selectedSights}
\end{figure*}

Figures~\ref{fig:smac-allResults-DsrVsOrigianl-win}, 
\ref{fig:smac-allResults-DsrVsOrigianl-Return}, 
\ref{fig:smac-allResults-MapVsStateVsPures-win}, and 
\ref{fig:smac-allResults-MapVsStateVsPures-Return} 
present the overall results for both test win rate and test return in SMAC.
Figures~\ref{fig:smac-allResults-5mVs6m}, 
\ref{fig:smac-allResults-8mVs9m}, 
\ref{fig:smac-allResults-10mVs11m}, 
\ref{fig:smac-allResults-3s_vs_5z}, 
\ref{fig:smac-allResults-3s5z}, and 
\ref{fig:smac-allResults-MMM2} 
display the detailed comparisons for each map in SMAC.
Figure~\ref{fig:smac-allResults-selectedSights} presents the full results of the selected sight ranges of DSR during testing within training in SMAC.

\subsection{Further SMAC-DT Results}

\begin{figure*}
    \centering
    \includegraphics[width=1\linewidth]{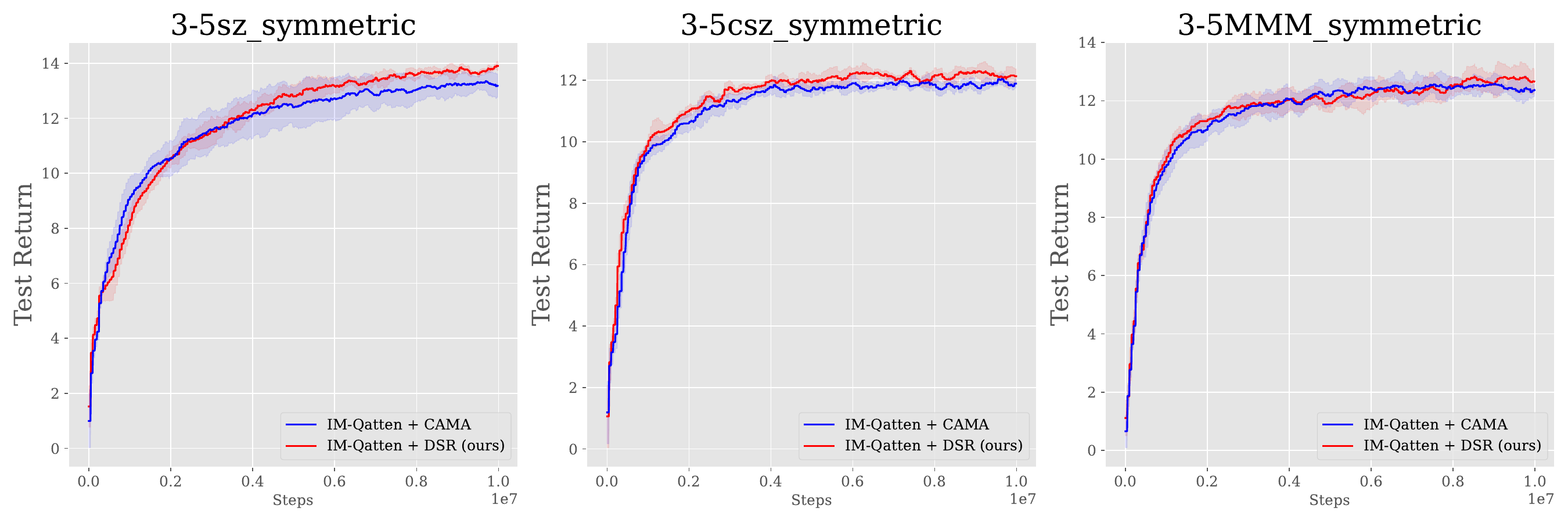}
    \caption{Test return in SMAC-DT.}
    \label{fig:smac-dt-allResults-return}
\end{figure*}

\begin{figure*}
    \centering
    \includegraphics[width=1\linewidth]{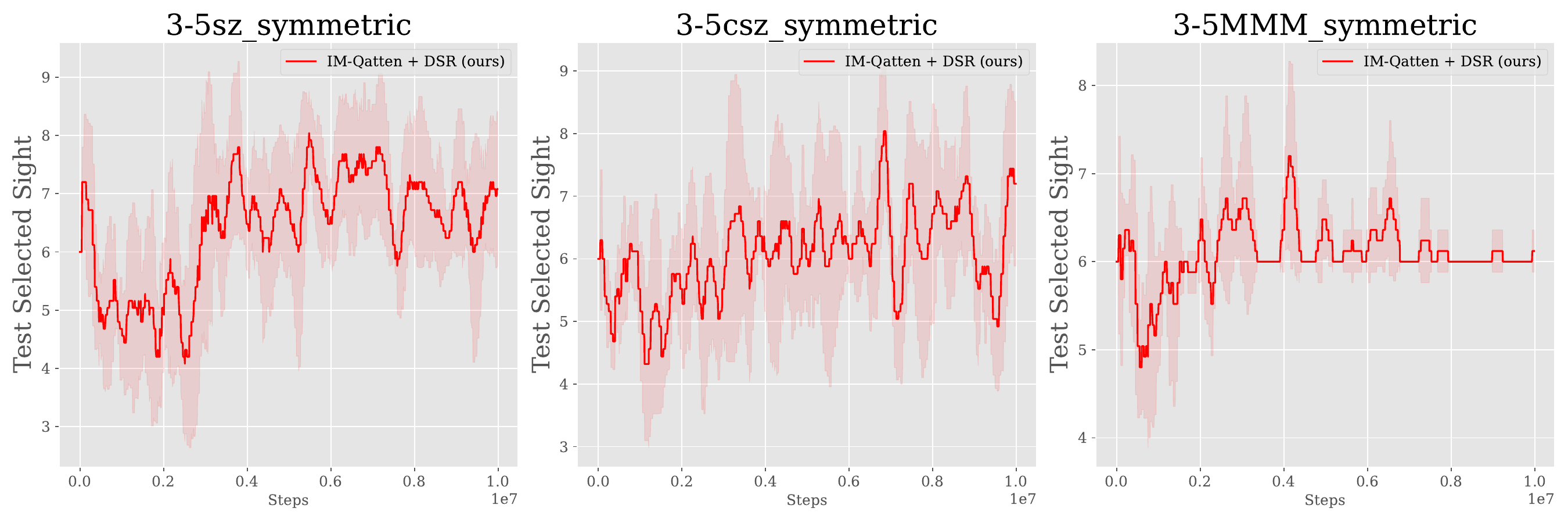}
    \caption{Test selected sights in SMAC-DT.}
    \label{fig:smac-dt-allResults-selectedSights}
\end{figure*}

We also report the test return (figure \ref{fig:smac-dt-allResults-return}) and provide the selected sight ranges by our DSR method during testing across the three tasks throughout the training process (figure \ref{fig:smac-dt-allResults-selectedSights}).

\subsection{More Algorithms on LBF and RWARE tasks}

\begin{figure*}
    \centering
    \includegraphics[width=1\linewidth]{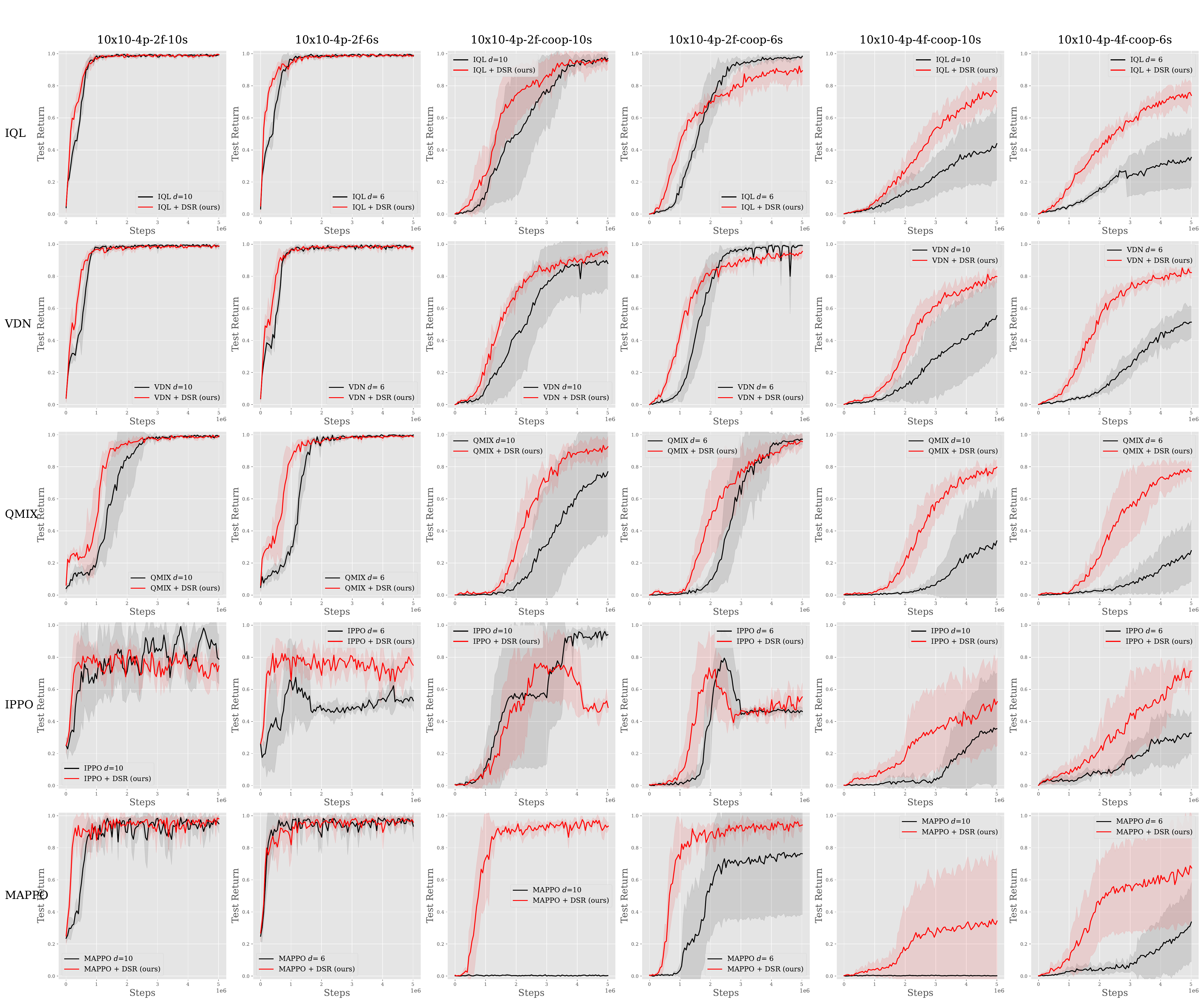}
    \caption{Test return of w/ and w/o DSR in different maps and original sight ranges in LBF.}
    \label{fig:lbf-allResults-DsrVsOriginal}
\end{figure*}

\begin{figure*}
    \centering
    \includegraphics[width=1\linewidth]{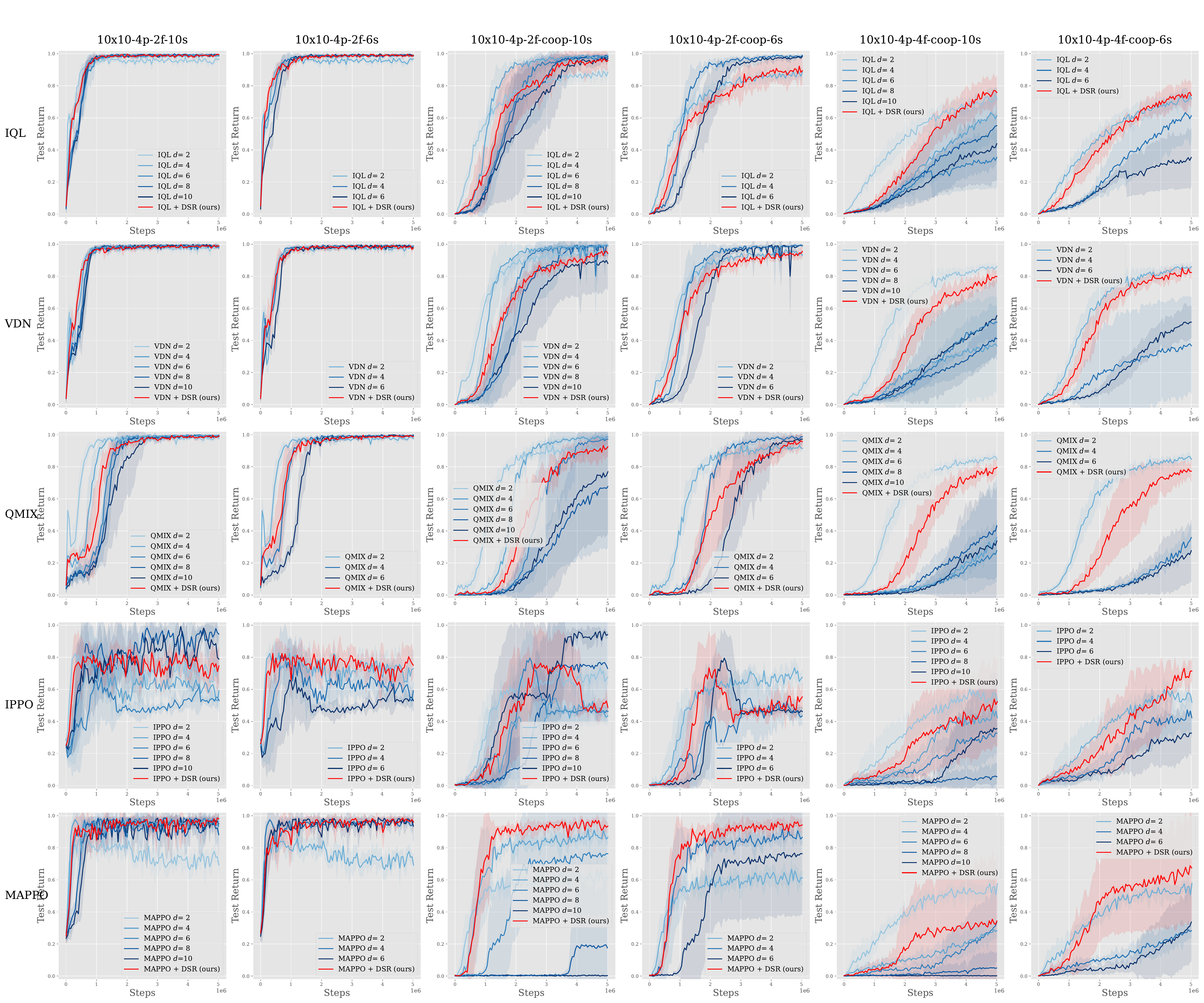}
    \caption{Test return across different maps and various sight ranges ($d$) in LBF. Note that the largest fixed sight range corresponds to the original sight range for that task (corresponding to the black line in Figure~\ref{fig:lbf-allResults-DsrVsOriginal}).}
    \label{fig:lbf-allResults-DsrVsPures}
\end{figure*}

\begin{figure*}
    \centering
    \includegraphics[width=1\linewidth]{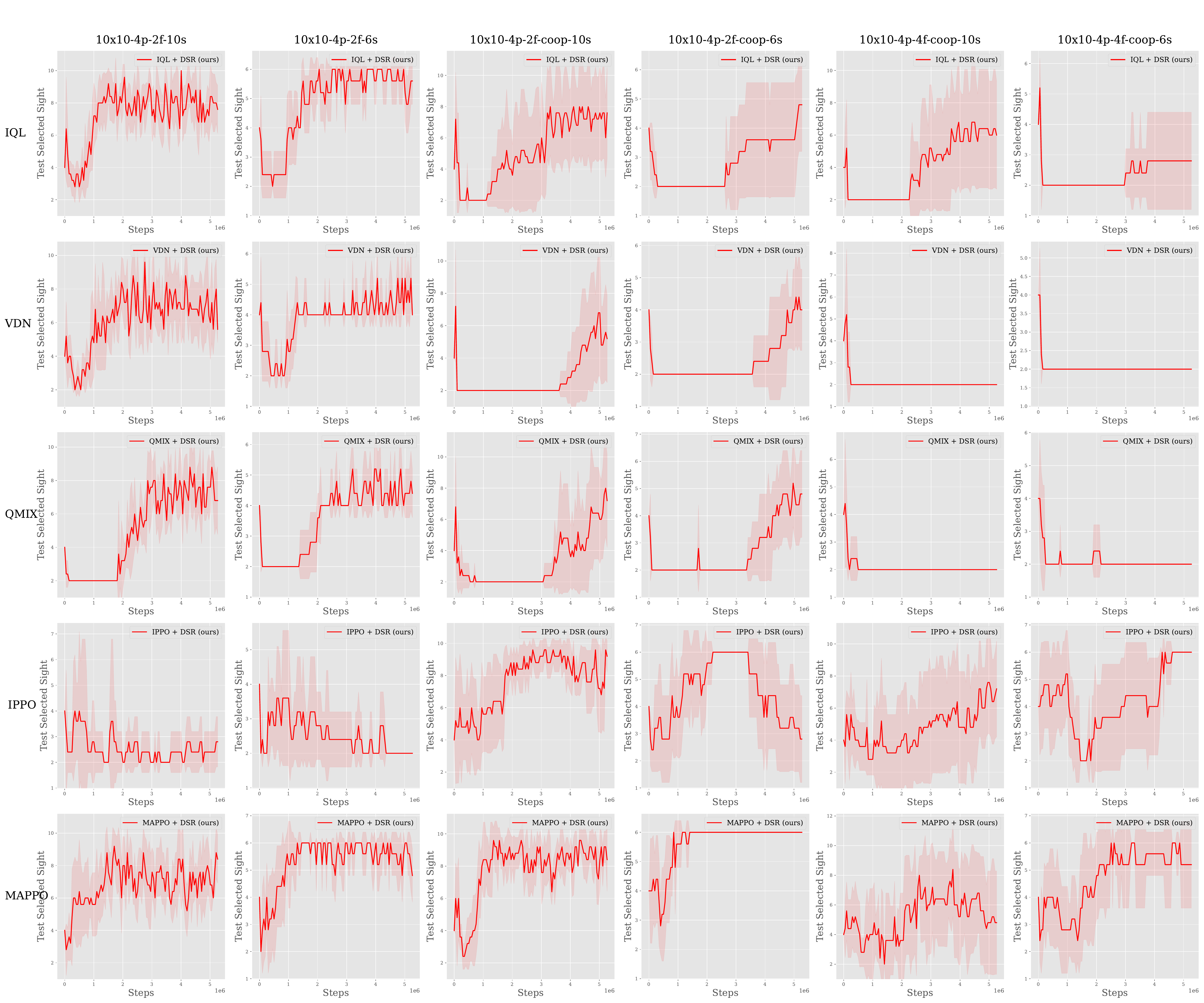}
    \caption{Test selected sights across different maps and various sight ranges in LBF.}
    \label{fig:lbf-allResults-selectedSights}
\end{figure*}

\begin{figure*}
    \centering
    \includegraphics[width=1\linewidth]{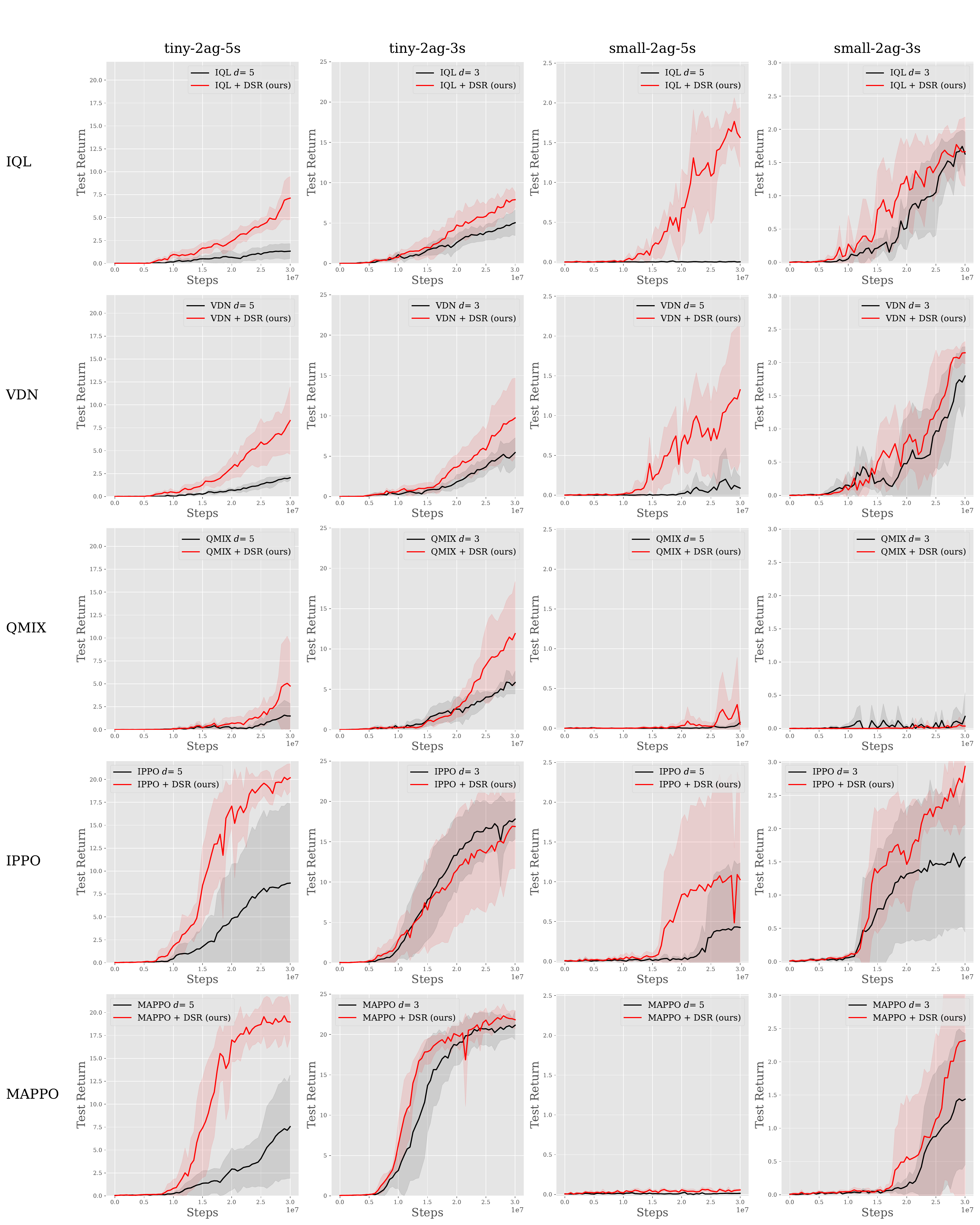}
    \caption{Test return of with DSR and without DSR in different maps and original sight ranges in RWARE.}
    \label{fig:rware-allResults-DsrVsOriginal}
\end{figure*}

\begin{figure*}
    \centering
    \includegraphics[width=1\linewidth]{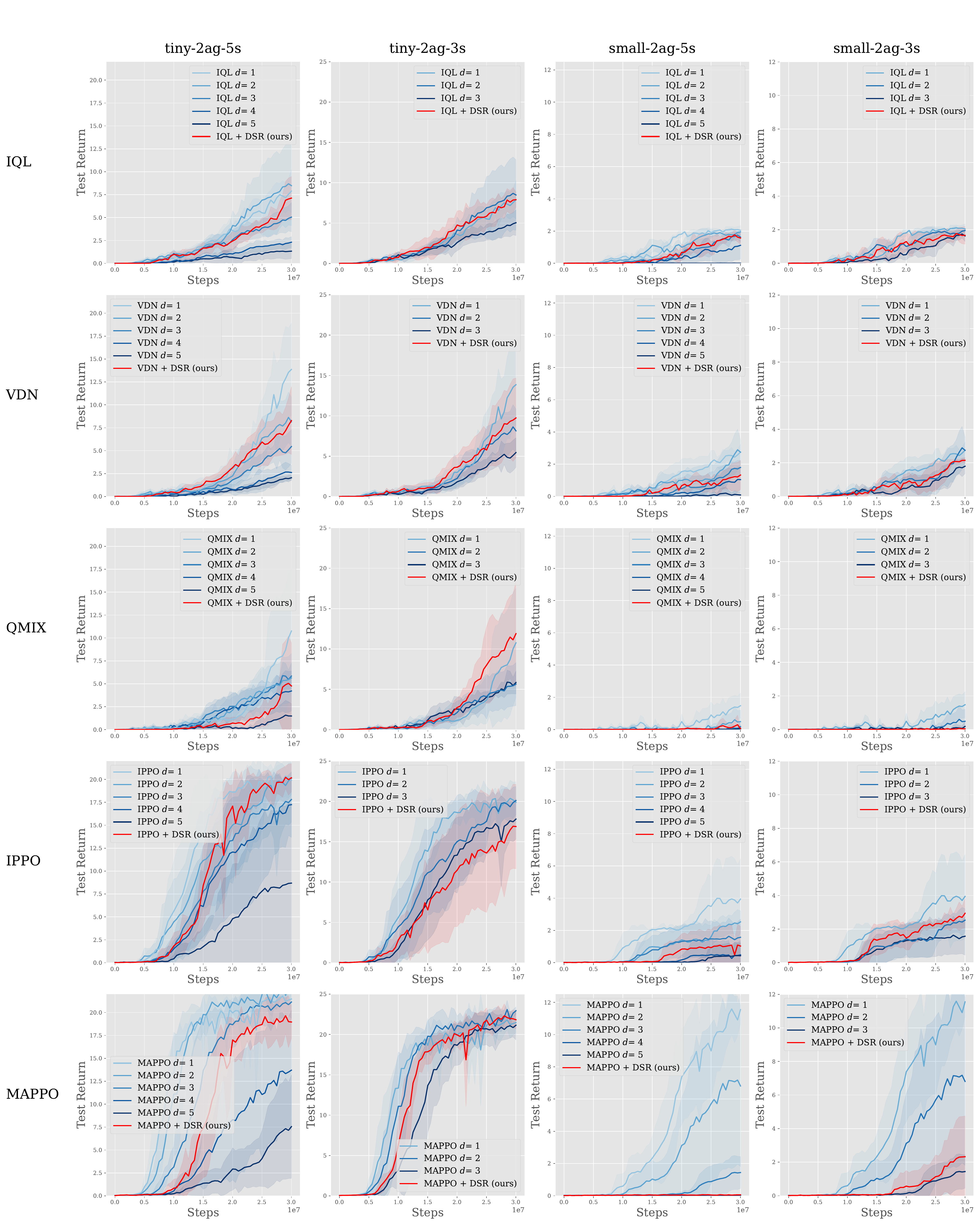}
    \caption{Test return across different maps and various sight ranges ($d$) in RWARE. Note that the largest fixed sight range corresponds to the original sight range for that task (corresponding to the black line in Figure~\ref{fig:rware-allResults-DsrVsOriginal}).}
    \label{fig:rware-allResults-DsrVsPures}
\end{figure*}

\begin{figure*}
    \centering
    \includegraphics[width=1\linewidth]{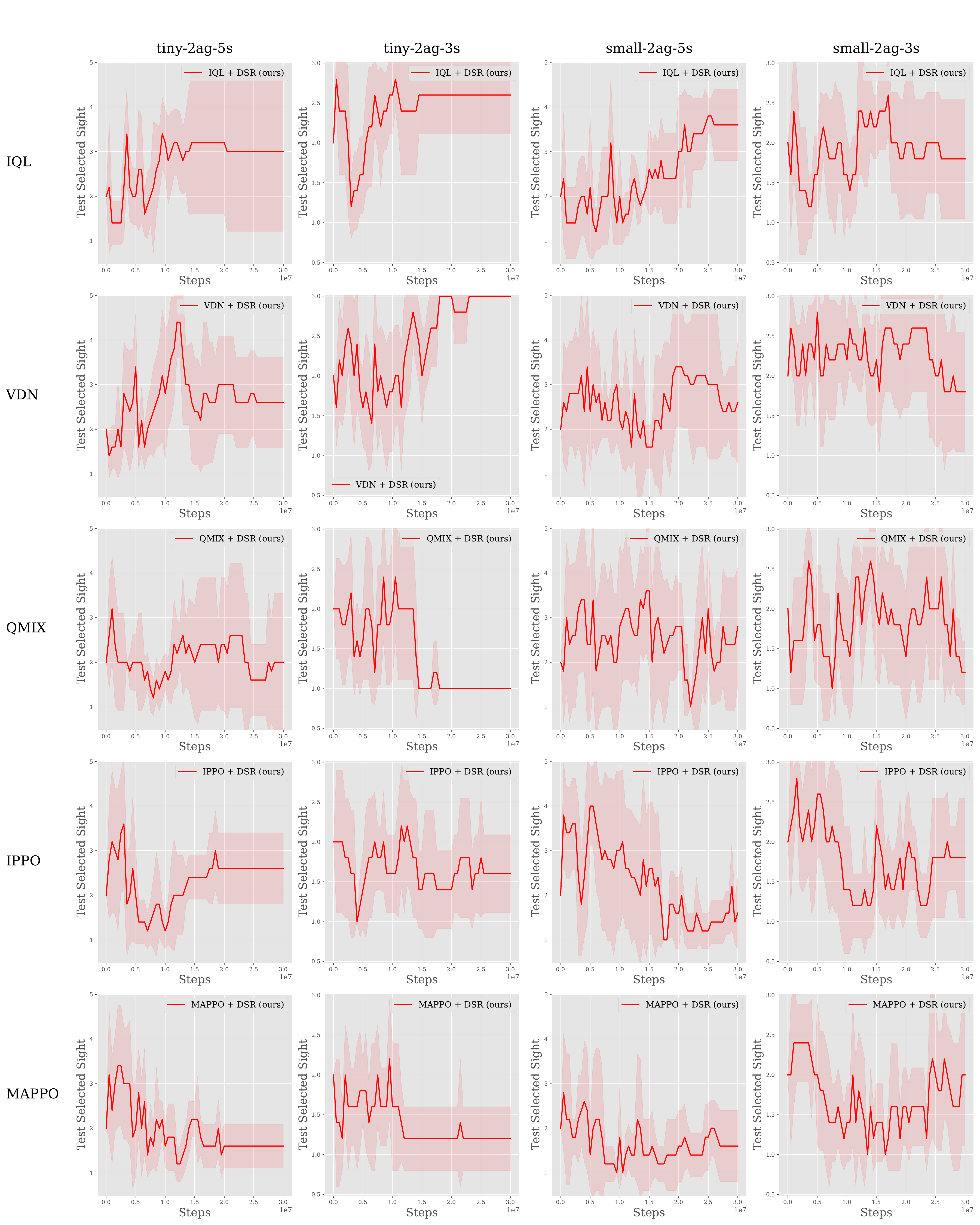}
    \caption{Test selected sights across different maps and various sight ranges in LBF.}
    \label{fig:rware-allResults-selectedSights}
\end{figure*}

Figures~\ref{fig:lbf-allResults-DsrVsOriginal}, \ref{fig:lbf-allResults-DsrVsPures}, \ref{fig:lbf-allResults-selectedSights}, \ref{fig:rware-allResults-DsrVsOriginal}, \ref{fig:rware-allResults-DsrVsPures} and \ref{fig:rware-allResults-selectedSights} present the results of IQL, VDN, IPPO, and MAPPO in LBF and RWARE. For easier comparison, the results of QMIX are also included.

\section{State Learning Complexity in SMAC}

\begin{figure}[htbp] 
    \centering
    \includegraphics[width=\linewidth]{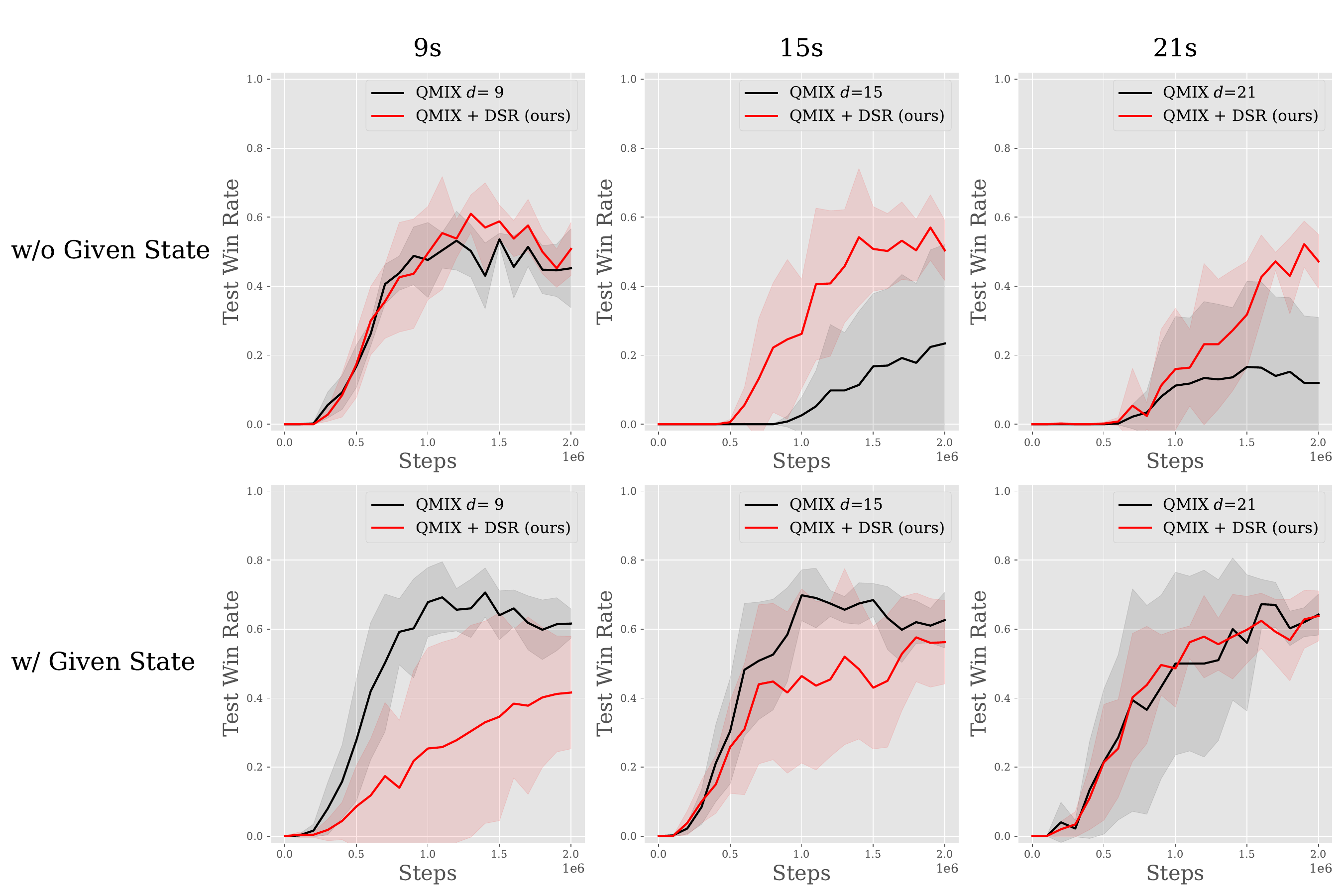}
    \caption{Mean test returns during training in SMAC 8m\_vs\_9m with different original sight ranges and state settings. Note that for the cases without the given state, the state input is formed by concatenating agents' observations.}
    \label{fig:smac-state-DsrVs} 
\end{figure}

In figure \ref{fig:smac-state-DsrVs}, we compare the impact of having full information through a provided global state in SMAC tasks. 
The global state used is provided by the environment itself. 
When the global state is not available, the input to the global value function is the concatenation of the agents' observations. 
We observe that, without a given global state, the performance of QMIX tends to decline as the sight range increases, whereas this trend is less noticeable when the global state is provided. 
In the absence of a given global state, since state features must be extracted from the observations, the sight range dilemma has a more pronounced effect.
As for the case with a given state, the finding that a larger sight range does not significantly affect the training results is somewhat similar to the results reported in \cite{yu_EnhancingMultiAgent_2023}.

\begin{figure}
    \centering
    \includegraphics[width=0.48\linewidth]{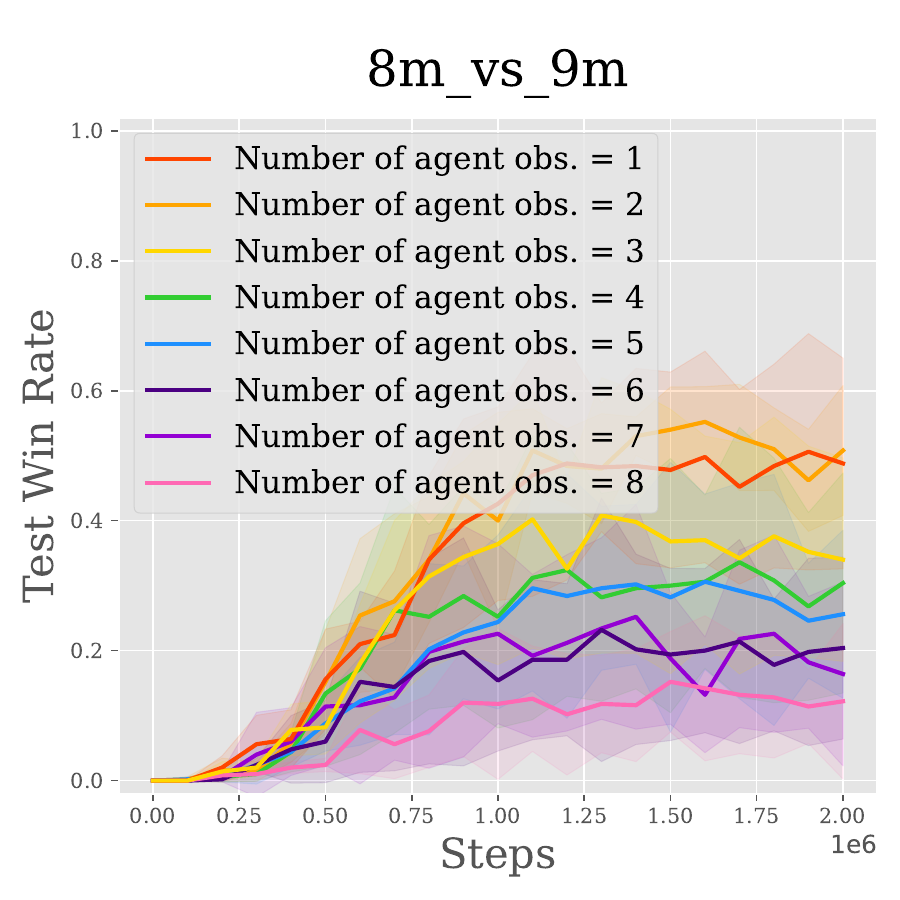}
    \caption{This map contains 8 agents (QMIX), and we compare the performance when combining different numbers of agent observations to form the state, with each agent having $d=\infty$.}
    \label{fig:smac-diff-num-ag-obs}
\end{figure}

We conducted an additional experiment to compare different configurations where the agents' sight ranges cover the entire map, and the state is composed of observations from various numbers of agents (in a fixed order).
As shown in figure \ref{fig:smac-diff-num-ag-obs}, the win rate decreases significantly as the number of agent observations used to form the state increases. 
This suggests that when there is overlap or redundancy in the agent observations, it becomes more difficult for the network to learn effectively. 
This finding implies that with a larger sight range, the increased likelihood of redundant information in the agent observations can hinder network training.
This also suggests that for methods heavily reliant on estimating the global state from observations, DSR could be a beneficial approach.

\section{Comparison between Using and Not Using GRU}

RNNs (e.g., GRU \cite{chung_EmpiricalEvaluation_2014}) are commonly used in MARL, especially in partially observable environments. 
We conducted an additional experiment comparing the performance of using GRU and not using GRU (i.e., using only MLP) on the LBF 10s-10x10-4p-4f-coop task. 
In this task, $d=10$ represents the original sight range, which allows the agent to observe the entire map.
Overall, it can be observed that GRU consistently outperforms MLP under the same $d$. In this task, both MLP and GRU achieve better performance with smaller sight ranges.

\begin{figure}
    \centering
    \includegraphics[width=0.75\linewidth]{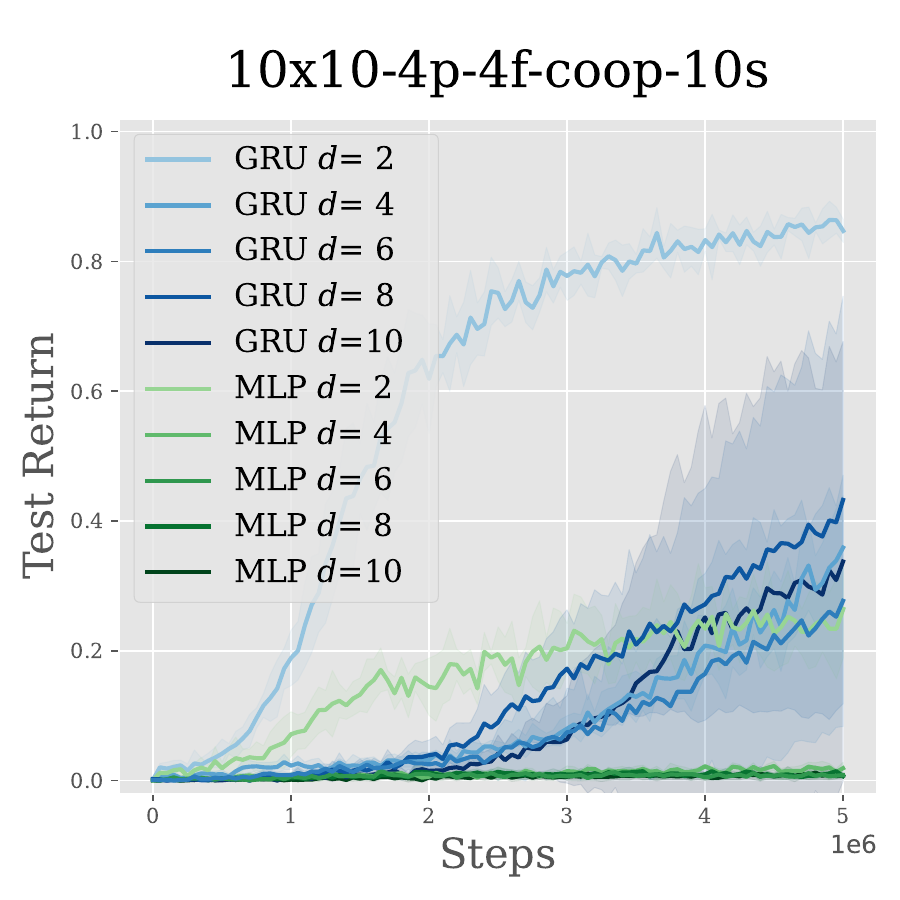}
    \caption{Comparison of QMIX results with varying fixed sight ranges on LBF 10s-10x10-4p-4f-coop, with and without RNNs.}
    \label{fig:enter-label}
\end{figure}

% -------------------------------------- LBF All Alg

\clearpage

\clearpage

\clearpage

% -------------------------------------- RWARE All Alg

\clearpage

\clearpage

\clearpage

% ------------------------------------------

% ------------------------------------------

\clearpage

\clearpage

% ---------------------------- SMAC-DT

\clearpage

%%%%%%%%%%%%%%%%%%%%%%%%%%%%%%%%%%%%%%%%%%%%%%%%%%%%%%%%%%%%%%%%%%%%%%%%

\end{document}